\documentclass{jpsj2}
\usepackage{txfonts}
\usepackage{bm}

\title{Momentum-Dependent Local Ansatz Approach to the Metallic Ferromagnetism}

\author{Yoshiro Kakehashi\thanks{yok@sci.u-ryukyu.ac.jp, to be published in J. Phys. Soc. Jpn.}}
\inst{Department of Physics, Faculty of Science,\\
University of the Ryukyus, \\
1 Senbaru, Nishihara, Okinawa, 903-0213, Japan} 

\abst{\vspace*{3mm}
The first principles momentum dependent local ansatz wavefunction method (MLA) has been extended to the ferromagnetic state by introducing spin-dependent variational parameters.  The theory is applied to the ferromagnetic Fe, Co, and Ni.  It is shown that the MLA yields the magnetizations being comparable to the results obtained by the GGA (generalized gradient approximation) in the density functional theory.  The projected momentum distribution functions as well as the mass enhancement factors are also calculated on the same footing, and are compared with those in the paramagnetic state.  It is shown that the calculated mass enhancement factor of Fe is strongly suppressed by the spin polarization due to exchange splitting of the e${}_{\rm g}$ flat bands, while those of Co and Ni remain unchanged by the polarization.  These results are shown to be consistent with the experimental results obtained from the low-temperature specific heats.
}

\kword{ first-principles theory, ferromagnetism, momentum-dependent local ansatz, Gutzwiller wevefunction, electron correlations, momentum distribution function, iron, cobalt, nickel}

\begin{document}
\maketitle

\section{Introduction}

The density functional theory (DFT) has played a central role in the quantitative description of the properties in solids over fifty years~\cite{slater72,fulde95,martin08}.  The theory is based on the Hohenberg-Kohn theorem~\cite{hohen64}, which guarantees that the ground-state energy is given by the functional of electron density, and  the Kohn-Sham scheme~\cite{kohn65} making use of the density of an independent electron system.  Using the exchange correlation potential obtained from the local density approximation~\cite{barth72} (LDA) or the generalized gradient approximation~\cite{perdew86} (GGA) for the electron gas model, the DFT quantitatively explained the stability of the structure, the lattice parameter, and the bulk modulus, as well as the magnetism of transition metals and 
compounds~\cite{janak76, janak77, moruzzi78}.

A large number of applications of the DFT to various systems, however, have also revealed significant problems and limitations of the theory.  For example, the accuracy of the DFT decreases with increasing Coulomb interaction strength as seen in the descriptions of the paramagnetism in $\epsilon$-Fe~\cite{pour14}, the weak antiferromagnetism in Fe-pnictides~\cite{imada10}, and the antiferromagnetism in cuprates~\cite{fulde12}.  The DFT is also limited to the ground-state properties connected to the charge and spin densities.  The excited states and excitation spectra cannot be described by the DFT since it is based on the Hohenberg-Kohn theorem.  For the same reason, the other physical quantities such as the charge and spin fluctuations described by the two-particle operators cannot be obtained by the DFT.  Furthermore, the momentum distribution function  (MDF) and related mass enhancement factor (MEF) cannot be calculated by the DFT because of the use of the Kohn-Sham independent electron scheme.

In order to remove the problems and limitations of the DFT mentioned above, various approaches have been developed.  One of the approaches is to construct the first principles tight-binding effective Hamiltonian~\cite{anis91,anis10,kake12} and to apply the wavefunction method~\cite{fulde95} to the Hamiltonian, in order to treat more correlated electron systems.  The first principles Gutzwiller wavefunction method~\cite{bune00,wang10,bune12,schick12} (GW) is a wellknown wavefunction approach for the ground-state electron correlations.  The GW~\cite{gutz63,gutz64,gutz65} describes the correlations controlling the amplitudes of atomic state at each site in solids.  The first principles GW theory clarified many physics of electron correlations in magnetism, the heavyfermion behaviors, and the metal-insulator transitions.

The Gutzwiller wavefunction, however, does not reduce to the second-order perturbation theory in the weak Coulomb interaction limit.  Therefore, it does not describe quantitatively the properties of correlated electron system.  In particular, the problem is serious for the description of the MEF associated with the low energy excitations in the vicinity of the Fermi surface, because it is obtained by a renormalization of the counterpart in the weak Coulomb interaction limit according to the Fermi liquid theory.

In order to overcome the difficulty of the GW, we recently proposed the first-principles momentum-dependent local ansatz (MLA) wavefunction~\cite{kake16,chan16,kake16-2,kake17}, which combines the MLA~\cite{kake08,pat11,pat13,pat13-2,kake14} with the first-principles tight-binding LMTO (linear muffin-tin orbital) LDA+U Hamiltonian~\cite{anis91,anis10,kake12}.  In the MLA, we first take into account the two particle excited states with momentum-dependent variational parameters in the momentum space, and project them onto the local orbitals.  The MLA overcomes the GW and reduces to the Rayleigh-Schr\"{o}dinger perturbation theory in the weak Coulomb interaction limit.

In our previous papers~\cite{kake16,chan16,kake16-2,kake17}, we investigated correlation effects in transition metals from Sc to Cu in the paramagnetic state.
We found strong suppression of charge fluctuations in Mn and Fe, obtained the Hund-rule coupling energies of Fe, Co, and Ni with the same order of magnitude as their Curie temperatures, as well as  the enhanced amplitudes of local moments in Mn and Fe being comparable to the effective Bohr magneton numbers~\cite{kake16-2}.  We also found that the MDF for $d$ electrons show a significant momentum dependence and a large deviation from the Fermi distribution function (FDF) due to the flat e${}_{\rm g}$ (t${}_{\rm 2g}$) bands on the Fermi level in Fe (Ni)~\cite{kake17}.

In the present paper, we extend the MLA to the spin-polarized case introducing the spin-dependent variational parameters as well as the self-consistent Hartree-Fock charge and exchange potentials.  We present the numerical results of the ground-state spin magnetizations and the projected momentum distribution functions as well as the mass enhancement factors in Fe, fcc Co, and Ni.
We demonstrate on the same footing of the MLA wavefunction that the first-principles MLA yields the magnetizations of Fe, Co, and Ni being comparable to the results of the GGA in the DFT, and that calculated MEF are consistent with the experimental data, so that the present approach is useful for the quantitative description of magnetism in correlated electron systems.

In the following section, we outline the spin-polarized MLA.  Introducing spin-dependent variational parameters, we derive the self-consistent equations as well as the expressions of the MDF in the ferromagnetic state.  In Sect. 3, we present the numerical results of the spin magnetizations and the projected MDF as well as the MEF.  We will show that calculated magnetizations for Co and Ni are in good agreement with the experiments, while the result for Fe is somewhat larger than the experimental value.  We will also show that the MEF of Fe is strongly suppressed by spin polarization due to exchange splitting of the e${}_{\rm g}$ bands, while the MEF for Co and Ni remain unchanged.  These results explain the experimental results obtained by the specific heat data.
In Sect. 4, we summarize the present results of the magnetizations, the MDF, and the MEF in the ferromagnetic Fe, Co, and Ni, and discuss the remaining problems to be solved in the future investigations.

\section{First-Principles MLA to the Ferromagnetic Metals}

\subsection{Wavefunction and variational principle}

We adopt the first-principles tight-binding LMTO LDA+U Hamiltonian with an atom per unit cell~\cite{anis10,ander94,anisimov97-2,anisimov93,kake08-2}. 
\begin{align}
H & = \sum_{iL\sigma}\epsilon^{0}_{L} \ n_{iL\sigma} 
+ \sum_{iLjL^{'}\sigma}{t}_{iLjL^{'}}\ 
a^{\dagger}_{iL\sigma}\,{a}_{jL^{'}\sigma}  \nonumber \\
  & \hspace*{-4mm} + \sum_{i}\Big[ 
\sum_{m} {U}_{mm}  n_{ilm\uparrow}\, n_{ilm\downarrow} 
+ \! \sum_{(m,m')} \!\! \Big(U_{mm'}-\frac{1}{2}J_{mm'}\Big) \,
n_{ilm} n_{ilm{'}} - 
 2 \!\! \sum_{(m,m')} \!\! J_{mm'}\,
\boldsymbol{s}_{ilm} \! \cdot \! \boldsymbol{s}_{ilm'} \Big] \,.
\label{eqhldau}
\end{align}
Here $\epsilon^{0}_{L}$ is an atomic level of orbital $ L$ on site $i$,  
${t}_{iLjL'}$  is a transfer integral between $iL$ and $jL'$,  
$L=(l, m)$ denotes the $s \,(l=0)$, $p\, (l=1)$, and $d \,(l=2)$ 
orbitals. 
$a^{\dagger}_{iL\sigma}{({a}_{iL\sigma})}$ is the creation (annihilation) 
operator for an electron on site $i$ with orbital $L$ and spin 
${\sigma}$, and 
$n_{iL\sigma} = a^{\dagger}_{iL\sigma}{a}_{iL\sigma}$ is 
the number operator on site $i$ with orbital $L$ and 
spin $\sigma$.  

The third term at the rhs (right-hand-side) of Eq. (\ref{eqhldau})
denotes the on-site Coulomb interactions between $d$ electrons.
$U_{mm}\,(U_{mm'} )$ and $J_{mm'}$ are the intra-orbital (inter-orbital) 
Coulomb and exchange interactions between $d$ electrons, respectively. 
$n_{ilm}\, (\boldsymbol{s}_{ilm})$ with $l = 2 $ is the charge 
(spin) density operator for $d$ electrons on site $i$ and orbital $m$. 

In the momentum-dependent local ansatz approach~\cite{chan16} (MLA), we rewrite the Hamiltonian $H$ as the sum of the Hartree-Fock Hamiltonian $H_{0}$ and the residual interactions $H_{\mathrm {I}}$: 
\begin{align}
H=H_{0}+H_{\mathrm{I}}\,.
\label{eqhhfi}
\end{align}
The latter is given by
\begin{align}
H_{\mathrm{I}}&= \sum_{i}{\Big[\sum_{L}U_{LL}^{(0)}\ {O}^{(0)}_{iLL}+\sum_{(L,L')}U_{LL'}^{(1)} \ {O}^{(1)}_{iLL'}+\sum_{(L,L')}U_{LL'}^{(2)}\ {O}^{(2)}_{iLL'}\Big]}\,.  
\label{eqhi}
\end{align}
The first term denotes the intra-orbital interactions, the second term is the inter-orbital charge-charge   interactions, and the third term expresses the inter-orbital spin-spin interactions, respectively. The Coulomb interaction energy parameters $U_{LL'}^{(\alpha)}$ are defined by $U_{LL}\delta_{LL'}$ $(\alpha=0)$, $U_{LL'}-J_{LL'}/2$ $(\alpha=1) $, and $-2J_{LL'}$ $ (\alpha=2)$, respectively. The two-particle operators ${O}^{(0)}_{iLL}$, ${O}^{(1)}_{iLL'}$, and ${O}^{(2)}_{iLL'}$  are defined by
\begin{equation}
{O}^{(\alpha)}_{iLL^{\prime}} = 
\begin{cases}
		 \ \delta n_{ilm\uparrow} \, 
\delta n_{ilm\downarrow} \, \delta_{LL^{\prime}} &  \ (\alpha=0) \\
\ \delta n_{ilm} \, \delta n_{ilm'}  &  \ (\alpha=1) \\ 
\ \delta \boldsymbol{s}_{ilm} \cdot 
\delta \boldsymbol{s}_{ilm'} & \ (\alpha=2)\, .
\end{cases}
\label{eqoalph}
\end{equation}
Note that $\delta A$ for an operator $A$ is defined by $\delta A=A-\langle A\rangle_{0}$, $\langle \sim \rangle_{0}$ being the average in the Hartree-Fock approximation. 

When the Hamiltonian $H$ is applied to the Hartree-Fock wavefunction
$|\phi \rangle$, the Hilbert space is expanded by the local operators 
$\{ O^{(\alpha)}_{iLL^{\prime}} \}$ in the residual interactions.
In order to take into account these states as well as the states
produced in the weak Coulomb interaction limit, we introduce the 
momentum-dependent local correlators 
$\lbrace\tilde{O}^{(\alpha)}_{iLL'}\rbrace$ ($\alpha=$ 0, 1, and 2) 
as follows.
\begin{align}
\tilde{O}^{(\alpha)}_{iLL'}&= \sum_{\{kn\sigma\}}\langle{k'_{2}n'_{2}\vert iL}\rangle_{\sigma'_{2}} \langle{iL\vert {k}_{2}{n}_{2}}\rangle_{\sigma_{2}} \langle{k'_{1}n'_{1}\vert iL'}\rangle_{\sigma'_{1}} \langle{iL'\vert {k}_{1}{n}_{1}}\rangle_{\sigma_{1}}  \nonumber \\
&\hspace{1cm}\times\lambda^{(\alpha)}_{{LL'}\{{2'2 1'1}\}}\ \delta(a^{\dagger}_{k'_{2}n'_{2}\sigma'_{2}}a_{{k}_{2}{n}_{2}\sigma_{2}})\ \delta(a^{\dagger}_{k'_{1}n'_{1}\sigma'_{1}}a_{{k}_{1}{n}_{1}\sigma_{1}})\,.
\label{eqotilde}
\end{align}
Here $a^{\dagger}_{k n \sigma }$ and $a_{k n\sigma}$ are the creation 
and annihilation operators for an electron with momentum $\bm{k}$, 
band index $n$, and spin $\sigma $. These operators are given by those 
in the site representation as 
$a_{k n\sigma}=\sum_{iL}a_{iL\sigma}\langle k n\vert iL\rangle_{\sigma}$\,. 
$\langle k n\vert iL\rangle_{\sigma}$ are the overlap integrals
between the Bloch state $(\bm{k}n\sigma)$ and the local-orbital state $(iL\sigma)$. 
$\lambda^{(\alpha)}_{{LL'}\{{2'2 1'1}\}}$ are the momentum-dependent variational parameters.
Note that the subscript $\{{2'2 1'1}\}$ is defined by 
$\{{2'2 1'1}\}=k^{\prime}_{2}n^{\prime}_{2}\sigma^{\prime}_{2}k_{2}n_{2}\sigma_{2}k^{\prime}_{1}n^{\prime}_{1}\sigma^{\prime}_{1}k_{1}n_{1}\sigma_{1}$

The two-particle correlators $\tilde{O}^{(0)}_{iLL}$, $\tilde{O}^{(1)}_{iLL'}$, and $\tilde{O}^{(2)}_{iLL'}$ describe the intra-orbital correlations, the inter-orbital charge-charge correlations, and the inter-orbital spin-spin correlations (, $i.e.,$ the Hund-rule correlations), respectively. Using the correlators $\lbrace\tilde{O}^{(\alpha)}_{iLL'}\rbrace$ and the Hartree-Fock ground-state wavefunction $\vert{\phi}\rangle$, we construct the MLA wavefunction as follows.
\begin{equation}
\vert{\Psi}_\mathrm{MLA}\rangle={\Big[\prod_{i}{\Big( 1-\sum_{L}{\tilde{O}}^{(0)}_{iLL}-\sum_{(L,L')}{\tilde{O}}^{(1)}_{iLL'}-\sum_{(L,L')}{\tilde{O}}^{(2)}_{iLL'}\Big) } \Big]}\,\ \vert{\phi}\rangle\,.
\label{mlawf}
\end{equation}

The variational parameters $\lambda^{(\alpha)}_{{LL'}\{{2'2 1'1}\}}$ in the correlators $\lbrace\tilde{O}^{(\alpha)}_{iLL'}\rbrace$ are determined by the variational principle for the ground-state energy $E$.
\begin{align}
\langle H\rangle=\langle H\rangle_{0}+N\epsilon_c \geq E \,.
\label{hvarp}
\end{align}
Here $\epsilon_c$ is the correlation energy per atom defined by $N\epsilon_c = \langle H \rangle-\langle H \rangle_{0}$, $N$ is the number of atoms, and $\langle \sim \rangle$ denotes the full average with respect to $\vert \Psi_\mathrm{MLA}\rangle$.  The correlation energy $\epsilon_c$ is given in the single-site approximation (SSA) as follows~\cite{chan16}. 
\begin{equation}
{\epsilon_c} =\frac{{- \langle {\tilde{O_i}^\dagger}} {H}_{I}\rangle_0 -\langle {H}_{I} \tilde{O_i}\rangle_0 
+\langle {\tilde{O_i}^\dagger} (\delta H) \tilde{O_i}\rangle_0 }{1+\langle{\tilde{O_i}^\dagger\tilde{O_i}}\rangle_0}\,. 
\label{corre}
\end{equation}
Here $\delta H = H - \langle H \rangle_{0}$, and the operator $\tilde{O_i}$ is defined by $\tilde{O_i}=\sum_{L}{\tilde{O}}^{(0)}_{iLL}+\sum_{(L,L')}{\tilde{O}}^{(1)}_{iLL'}+\sum_{(L,L')}{\tilde{O}}^{(2)}_{iLL'}$.

\subsection{Self-consistent equations for variational parameters}

In order to simplify calculations, we adopt the following ansatz for the variational parameters, which interpolates between the weak Coulomb interaction limit and the atomic limit~\cite{chan16,kake17}. 
\begin{equation}
\lambda^{(\alpha)}_{{LL'}\{{2'2 1'1}\}}=\frac{U_{LL'}^{(\alpha)}\sum_{\tau}C_{\tau\sigma_{2}\sigma_{2}^{'}\sigma_{1}\sigma_{1}^{'}}^{(\alpha)}\ \tilde{\lambda}_{\alpha\tau L L'}^{(\sigma_{2}\sigma_{1})}}{\Delta E_{k'_{2}n'_{2}\sigma'_{2} k_{2}n_{2}\sigma_{2} k'_{1}n'_{1}\sigma'_{1} k_{1}n_{1}\sigma_{1}}-\epsilon_c}\,.
\label{lambda1}
\end{equation}
Here $\Delta E_{k'_{2}n'_{2}\sigma'_{2} k_{2}n_{2}\sigma_{2} k'_{1}n'_{1}\sigma'_{1} k_{1}n_{1}\sigma_{1}}$ is the two-particle excitation energy defined by $\Delta E_{k'_{2}n'_{2}\sigma'_{2} k_{2}n_{2}\sigma_{2} k'_{1}n'_{1}\sigma'_{1} k_{1}n_{1}\sigma_{1}}=\epsilon_{k'_{2}n'_{2}\sigma'_{2}}-\epsilon_{k_{2}{n}_{2}\sigma_{2}}+\epsilon_{k'_{1}n'_{1}\sigma'_{1}}-\epsilon_{k_{1}{n}_{1}\sigma_{1}}$.
$\epsilon_{kn\sigma}$ denotes the Hartree-Fock one
electron energy eigenvalue with momentum $\boldsymbol{k}$, 
band index $n$, and spin $\sigma$.
The spin-dependent coefficients 
$C_{\tau\sigma_{2}\sigma_{2}^{'}\sigma_{1}\sigma_{1}^{'}}^{(\alpha)}$ 
are defined by 
$\delta_{\sigma'_{2}\downarrow}\,\delta_{\sigma_{2}\downarrow}\,
\delta_{\sigma'_{1}\uparrow}\,\delta_{\sigma_{1}\uparrow}$  
($\alpha=0$), 
$\delta_{\sigma'_{2}\sigma_{2}} \,\delta_{\sigma'_{1}\sigma_{1}}$  
($\alpha=1$), 
$-(1/4)\ \ \sigma_{1}\sigma_{2}\delta_{\sigma'_{2}\sigma_{2}} 
\delta_{\sigma'_{1}\sigma_{1}}$ ($\alpha=2,\tau=l$), and 
$-(1/2)\sum_{\sigma}\delta_{\sigma'_{2} - \sigma} 
\delta_{\sigma_{2}\sigma} \delta_{\sigma'_{1}\sigma} 
\delta_{\sigma_{1}-\sigma}$ ($\alpha=2,\tau=t$), respectively. 
Note that $l\,(t)$ implies the longitudinal (transverse) component. 

In the ferromagnetic state, the renormalization factors 
$\tilde{\lambda}_{\alpha\tau L L'}^{(\sigma\sigma')}$ in Eq. (\ref{lambda1}) have a form as
\begin{equation}
\tilde{\lambda}_{\alpha\tau L L'}^{(\sigma\sigma')} = 
\begin{cases}
		 \ \tilde{\lambda}_{0LL} \delta_{LL'}\delta_{\sigma'-\sigma} &  \ (\alpha=0) \\
\ \tilde{\lambda}_{1LL'} +  \tilde{\lambda}^{(s)}_{1LL'} \sigma \, \delta_{\sigma' -\sigma} &  \ (\alpha=1) \\ 
\ \tilde{\lambda}_{2lLL'} + \tilde{\lambda}^{(s)}_{2lLL'} \sigma \, \delta_{\sigma' \sigma} & \ (\alpha=2,\tau=l) \\
\ \big( \tilde{\lambda}_{2tLL'} + \tilde{\lambda}^{(s)}_{2tLL'} \sigma \big) 
\delta_{\sigma' -\sigma} & \ (\alpha=2,\tau=t)  \, .
\end{cases}
\label{lambda2}
\end{equation}
Note that when $\tilde{\lambda}_{0LL}=\tilde{\lambda}_{1LL'}=1$ and $\tilde{\lambda}_{2lLL'}=\tilde{\lambda}_{2tLL'}=-1$, and $\tilde{\lambda}^{(s)}_{1LL'}=\tilde{\lambda}^{(s)}_{2lLL'}=\tilde{\lambda}^{(s)}_{2tLL'}=0$, the MLA wavefunction (\ref{mlawf})
reduces to that of the Rayleigh-Schr\"{o}dinger perturbation theory in
the weak Coulomb interaction limit.
The renormalization factors $\{ \tilde{\lambda}_{\alpha\tau L L'}^{(\sigma\sigma')} \}$ are the new variational parameters to be determined.

Substituting Eq. (\ref{lambda1}) into the elements in Eq. (\ref{corre}), we obtain the following forms.
\begin{equation}
\langle {H}_{I} \tilde{O_i}\rangle_0=\sum_{\alpha\alpha'}\sum_{<LL'>}\sum_{<L''L'''>}U_{LL'}^{(\alpha)}\ U_{L''L'''}^{(\alpha')}\sum_{\tau\sigma\sigma'}\tilde{\lambda}_{\alpha'\tau L'' L'''}^{(\sigma\sigma')}\ P_{\tau LL'L'' L'''\sigma\sigma'}^{(\alpha\alpha')}\,,
\label{equ27}
\end{equation}
\begin{align}
\langle {\tilde{O_i}^\dagger} \! (\delta{H}_{0}) \tilde{O_i}\rangle_0=\sum_{\alpha\alpha'}\sum_{<LL'>}\sum_{<L''L'''>}\!\!U_{LL'}^{(\alpha)}\ U_{L''L'''}^{(\alpha')}
\sum_{\tau\sigma\sigma'}\sum_{\tau'\sigma''\sigma'''}
\!\tilde{\lambda}_{\alpha\tau L L'}^{(\sigma\sigma')} \,\tilde{\lambda}_{\alpha'\tau' L'' L'''}^{(\sigma''\sigma''')}\,Q_{\tau\tau' LL'L'' L'''\sigma\sigma'\sigma''\sigma'''}^{(\alpha\alpha')}\,,
\label{equ28}
\end{align}
\begin{equation}
\langle {\tilde{O_i}^\dagger} {H}_{I} \tilde{O_i}\rangle_0=\sum_{\alpha}\sum_{<LL'>}U_{LL'}^{(\alpha)} \sum_{\tau\sigma\sigma'} \tilde{\lambda}_{\alpha\tau L L'}^{(\sigma\sigma')}\  K_{\tau L L'\sigma\sigma'}^{(\alpha)}\,,
\label{equ29}
\end{equation}
\begin{align}
\langle {\tilde{O_i}^\dagger} \tilde{O_i}\rangle_0=\sum_{\alpha\alpha'}\sum_{<LL'>}\sum_{<L''L'''>}U_{LL'}^{(\alpha)}\ U_{L''L'''}^{(\alpha')}
\sum_{\tau\sigma\sigma'}\sum_{\tau'\sigma''\sigma'''} \tilde{\lambda}_{\alpha\tau L L'}^{(\sigma\sigma')}\ \tilde{\lambda}_{\alpha'\tau' L'' L'''}^{(\sigma''\sigma''')}\ S_{\tau\tau' LL'L'' L'''\sigma\sigma'\sigma''\sigma'''}^{(\alpha\alpha')}\,.
\label{equ31}
\end{align}
Here the sum $\sum_{<LL'>}$ is defined by $\sum_{L}$ for $L' = L$ and $\sum_{(L,L')}$ for 
$L' \ne L$.  The coefficients $P_{\tau LL'L'' L'''\sigma\sigma'}^{(\alpha\alpha')}, Q_{\tau \tau'LL'L'' L'''\sigma\sigma'\sigma''\sigma'''}^{(\alpha\alpha')}$, and $S_{\tau\tau'LL'L'' L'''\sigma\sigma'\sigma''\sigma'''}^{(\alpha\alpha')}$ are calculated by making use of Wick's theorem and the Laplace transformations. $K_{\tau L L'\sigma\sigma'}^{(\alpha)}$ in Eq. (\ref{equ29}) are the higher order corrections in the Coulomb interactions $\{U_{LL'}^{(\alpha)}\}$. They have the following form.
\begin{equation}
K_{\tau L L'\sigma\sigma'}^{(\alpha)}=\sum_{\alpha'}\sum_{<L''L'''>}\sum_{\tau'\sigma''\sigma'''}U_{L''L'''}^{(\alpha')}\ R_{\tau\tau' LL'L'' L'''\sigma\sigma'\sigma''\sigma'''}^{(\alpha\alpha')}\ \tilde{\lambda}_{\alpha'\tau' L'' L'''}^{(\sigma''\sigma''')}\,.
\label{equ30}
\end{equation}
The coefficients $R_{\tau\tau' LL'L'' L'''\sigma\sigma'\sigma''\sigma'''}^{(\alpha\alpha')}$ can be calculated again with use of Wick's theorem.

The self-consistent equations for the variational parameters $\tilde{\lambda}_{\alpha\tau L L'}^{(\sigma\sigma')}$ are obtained from the variational principle (\ref{hvarp}) as follows. 
\begin{align}
&\sum_{\alpha'}\sum_{<L''L'''>}\sum_{\tau' \sigma''\sigma'''}U_{L''L'''}^{(\alpha')}
\left( Q_{\tau\tau' LL'L'' L'''\sigma\sigma'\sigma''\sigma'''}^{(\alpha\alpha')} - 
\epsilon_c \ S_{\tau\tau' LL'L'' L'''\sigma\sigma'\sigma''\sigma'''}^{(\alpha\alpha')} \right)\ \tilde{\lambda}_{\alpha'\tau' L'' L'''}^{(\sigma''\sigma''')} \nonumber \\
&\hspace{5cm} = \sum_{\alpha'}\sum_{<L''L'''>} U_{L''L'''}^{(\alpha')}\ P_{\tau L'' L'''LL'\sigma\sigma'}^{(\alpha'\alpha)} - K_{\tau L L'\sigma\sigma'}^{(\alpha)} \,.
\label{eqvp1}
\end{align}
Since we can verify the relations $Q_{\tau\tau' LL'L'' L'''\sigma\sigma'\sigma''\sigma'''}^{(\alpha\alpha')} \varpropto \delta_{\tau\tau^{\prime}} \delta _{<LL^{\prime}><L''L'''>} \delta_{\sigma^{\prime\prime}\sigma} \delta_{\sigma^{\prime\prime\prime}\sigma^{\prime}}$, 
$S_{\tau\tau' LL'L'' L'''\sigma\sigma'\sigma''\sigma'''}^{(\alpha\alpha')} \varpropto \delta_{\tau\tau^{\prime}} \delta _{<LL^{\prime}><L''L'''>} \delta_{\sigma^{\prime\prime}\sigma} \delta_{\sigma^{\prime\prime\prime}\sigma^{\prime}}$, and $P_{\tau L'' L'''LL'\sigma\sigma'}^{(\alpha'\alpha)} \varpropto \delta _{<LL^{\prime}><L''L'''>}$, the above equations reduce to
\begin{align}
&\sum_{\alpha'} U_{LL'}^{(\alpha')} \, 
\tilde{Q}_{\tau LL' \sigma\sigma'}^{(\alpha\alpha')} \tilde{\lambda}_{\alpha'\tau L L'}^{(\sigma\sigma')}  = 
\sum_{\alpha'} U_{LL'}^{(\alpha')} \, P_{\tau LL' \sigma\sigma'}^{(\alpha'\alpha)} - K_{\tau LL' \sigma\sigma'}^{(\alpha)} \,.
\label{eqvp2}
\end{align}
Here $\tilde{Q}_{\tau LL' \sigma\sigma'}^{(\alpha\alpha')} = Q_{\tau\tau LL'L L' \sigma\sigma'\sigma\sigma'}^{(\alpha\alpha')} - \epsilon_c \ S_{\tau\tau LL'L L' \sigma\sigma'\sigma\sigma'}^{(\alpha\alpha')}$ and
$P_{\tau LL' \sigma\sigma'}^{(\alpha'\alpha)} = P_{\tau LL'LL'\sigma\sigma'}^{(\alpha'\alpha)}$.  
The matrix elements of $\tilde{Q}_{\tau LL' \sigma\sigma'}^{(\alpha\alpha')}$, $P_{\tau LL' \sigma\sigma'}^{(\alpha'\alpha)}$, and $K_{\tau LL' \sigma\sigma'}^{(\alpha)}$ are given in Appendix A.

It should be noted that the charge and exchange potentials in the atomic levels $\epsilon_{iL\sigma}$ of the Hartree-Fock wavefunction $|\phi \rangle$ can also be treated as variational parameters.
\begin{equation}
\epsilon_{iL\sigma} = \epsilon^{0}_{L} + \Big( U^{(0)}_{LL} \bar{n}_{L -\sigma} + \sum_{L^{\prime} \neq L} U^{(1)}_{LL'} \, \bar{n}_{L'} + \frac{1}{4} \sum_{L^{\prime} \neq L} U^{(2)}_{LL'} \, \bar{m}_{L'} \sigma \Big) \, \delta_{l2} \,.
\label{vare}
\end{equation}
Here $\bar{n}_{L}$ and $\bar{m}_{L}$ are the trial charge and magnetization parameters for electrons of the orbital $L$, and $\bar{n}_{L\sigma}=(\bar{n}_{L} + \sigma \bar{m}_{L})/2$ denotes the trial electron number of orbital $L$ and spin $\sigma$.
We adopt here the following ansatz for simplicity without taking further variations.
\begin{equation}
\bar{n}_{L} = \langle n_{iL} \rangle \, ,
\label{ansatzn}
\end{equation}
\begin{equation}
\bar{m}_{L} = \langle m_{iL} \rangle \, .
\label{ansatzm}
\end{equation}

The partial electron number and magnetization of orbital $L$ on site $i$ are given as follows in the SSA. 
\begin{equation}
\langle n_{iL} \rangle = \langle n_{iL}\rangle_{0} + \frac{\langle\tilde{O}_{i}^{\dagger} (\delta{n}_{iL})\tilde{O}_{i}\rangle_{0}}{1+\langle{\tilde{O_i}^\dagger\tilde{O_i}}\rangle_0} \, ,
\label{nil}
\end{equation}
\begin{equation}
\langle m_{iL} \rangle = \langle m_{iL}\rangle_{0} + \frac{\langle\tilde{O}_{i}^{\dagger} (\delta{m}_{iL})\tilde{O}_{i}\rangle_{0}}{1+\langle{\tilde{O_i}^\dagger\tilde{O_i}}\rangle_0} \, .
\label{mil}
\end{equation}
Here $\langle n_{iL}\rangle_{0}$ ($\langle m_{iL} \rangle_{0}$) denotes the average electron number (magnetization) with respect to the Hartree-Fock wavefunction. The second terms at the right-hand-side (rhs) are the correlation corrections.
Explicit expressions for $\langle\tilde{O}_{i}^{\dagger}(\delta{n}_{iL})\tilde{O}_{i}\rangle_{0}$, $\langle\tilde{O}_{i}^{\dagger}(\delta{m}_{iL})\tilde{O}_{i}\rangle_{0}$, and 
$\langle{\tilde{O_i}^\dagger\tilde{O_i}}\rangle_0$ are given in Appendix B.

The Fermi level $\epsilon_{F}$ is determined from the condition for the conduction electron number $n_{e}$, 
\begin{equation}
n_{e}=\sum_{L}\langle n_{iL}\rangle\,.
\label{nefermi}
\end{equation}

Equation (\ref{corre}) for the correlation energy $\epsilon_{c}$, Eq. (\ref{eqvp2}) for the variational parameters $\{ \tilde{\lambda}_{\alpha\tau L L'}^{(\sigma\sigma')} \}$, Eqs. (\ref{ansatzn})-(\ref{mil}) for the self-consistent conditions for the charge and exchange potentials, and Eq. (\ref{nefermi}) for the Fermi level $\epsilon_{F}$ have to be solved self-consistently because they are coupled each other.

After we solve the coupled equations, we can calculate the magnetization per atom according to the following expression.
\begin{align}
\langle m_{i} \rangle = \sum_{L} \, \langle m_{iL} \rangle \,.
\label{milcorr}
\end{align}

\subsection{Momentum distribution function and mass enhancement factor}

The momentum distribution function (MDF) is given as follows.
\begin{equation}
\langle n_{kn\sigma}\rangle=f(\tilde{\epsilon}_{kn\sigma})+\frac{N\langle\tilde{O}_{i}^{\dagger}(\delta{n}_{kn\sigma}) \tilde{O}_{i}\rangle_{0}}{1+\langle{\tilde{O_i}^\dagger\tilde{O_i}}\rangle_0}\,.
\label{nknsigma}
\end{equation}
The first term at the rhs is the MDF for the Hartree-Fock independent electrons, {\it i.e.},   the Fermi distribution function (FDF) at zero temperature $f(\tilde{\epsilon}_{kn\sigma})$. $\tilde{\epsilon}_{kn\sigma}$ is the Hartree-Fock one-electron energy measured from the Fermi level $\epsilon_{F}$. The second term at the rhs of Eq. (\ref{nknsigma}) is the correlation correction.  The numerator $N\langle\tilde{O}_{i}^{\dagger}(\delta{n}_{kn\sigma}) \tilde{O}_{i}\rangle_{0}$ is given in Appendix C.

The quasiparticle weight $Z_{{k_{F}}n\sigma}$ is obtained from the jump of $\langle n_{kn\sigma}\rangle$ at the Fermi level $\epsilon_{F}$. 
Taking the average over the Fermi surface, we obtain the average quasiparticle weight $Z_{\sigma}$.
\begin{equation}
Z_{\sigma}=1+\frac{\overline{\delta(N\langle\tilde{O}_{i}^{\dagger}(\delta{n}_{kn\sigma}) \tilde{O}_{i}\rangle_{0})_{k_{F}}}}{1+\langle{\tilde{O_i}^\dagger\tilde{O_i}}\rangle_0}\,.
\label{zsigma}
\end{equation}
The second term at the rhs is the correlation corrections. The upper bar in the numerator denotes the average over the Fermi surface, and ${\delta(N\langle\tilde{O}_{i}^{\dagger}(\delta{n}_{kn\sigma}) \tilde{O}_{i}\rangle_{0})_{k_{F}}}$ means the jump at the wavevector $\bm{k}_{F}$ on the Fermi surface. 

The explicit expression of ${\overline{\delta(N\langle\tilde{O}_{i}^{\dagger}(\delta{n}_{kn\sigma}) \tilde{O}_{i}\rangle_{0})_{k_{F}}}}$ is given by
\begin{align}
\overline{\delta{(N\langle\tilde{O}_{i}^{\dagger} (\delta{n}_{kn\sigma}) \tilde{O}_{i}\rangle_{0})}}_{k_{F}} &=
- \sum_{L} {U_{LL}^{(0)}}^{2} \tilde{\lambda}_{0LL}^{2} 
\dfrac{\rho_{L\sigma}(\epsilon_{F})}{\rho_{\sigma}(\epsilon_{F})} 
(B_{LL -\sigma\sigma}(\epsilon_{F}) + C_{LL -\sigma\sigma}(\epsilon_{F})) \nonumber \\
&\hspace*{-25mm} - \sum_{(L,L')} \sum_{\sigma'} \Big[ \Big( \sigma U^{(1)}_{LL'} 
\tilde{\lambda}^{(\sigma'\sigma)}_{1LL'} - \frac{1}{4} \sigma' U^{(2)}_{LL'} 
\tilde{\lambda}^{(\sigma'\sigma)}_{2lLL'} \Big)^{2}
\dfrac{\rho_{L'\sigma}(\epsilon_{F})}{\rho_{\sigma}(\epsilon_{F})} 
(B_{LL' \sigma'\sigma}(\epsilon_{F}) + C_{LL' \sigma'\sigma}(\epsilon_{F}))  \nonumber \\
&\hspace*{-10mm} + \Big( \sigma U^{(1)}_{LL'} 
\tilde{\lambda}^{(\sigma\sigma')}_{1LL'} - \frac{1}{4} \sigma' U^{(2)}_{LL'} 
\tilde{\lambda}^{(\sigma\sigma')}_{2lLL'} \Big)^{2}
\dfrac{\rho_{L\sigma}(\epsilon_{F})}{\rho_{\sigma}(\epsilon_{F})} 
(B_{L'L \sigma'\sigma}(\epsilon_{F}) + C_{L'L \sigma'\sigma}(\epsilon_{F})) \Big]
\nonumber \\
&\hspace*{-30mm} - \frac{1}{4} \sum_{(L,L')}  {U^{(2)}_{LL'}}^{2} \Big[ 
\tilde{\lambda}^{(-\sigma \sigma) \, 2}_{2tLL'} \, \Big(  
\dfrac{\rho_{L\sigma}(\epsilon_{F})}{\rho_{\sigma}(\epsilon_{F})} 
B_{tLL' \sigma}(\epsilon_{F}) 
+ \dfrac{\rho_{L'\sigma}(\epsilon_{F})}{\rho_{\sigma}(\epsilon_{F})}  
C_{tLL' \sigma}(\epsilon_{F}) \Big)   \nonumber \\
&\hspace*{-8mm} + \tilde{\lambda}^{(\sigma -\sigma) \, 2}_{2tLL'} \, \Big(  
\dfrac{\rho_{L'\sigma}(\epsilon_{F})}{\rho_{\sigma}(\epsilon_{F})} 
B_{tL'L \sigma}(\epsilon_{F}) 
+ \dfrac{\rho_{L\sigma}(\epsilon_{F})}{\rho_{\sigma}(\epsilon_{F})}  
C_{tL'L \sigma}(\epsilon_{F}) \Big)
\Big] \, .
\label{avznum}
\end{align}
Here $\rho_{L\sigma}(\epsilon_{F})$ is the local density of states (LDOS) for electrons with orbital $L$ and spin $\sigma$ at $\epsilon_{F}$, $\rho_{\sigma}(\epsilon_{F})$ is the density of states (DOS) per site for electrons with spin $\sigma$ at $\epsilon_{F}$, and
\begin{align}
B_{LL' \sigma'\sigma}(\epsilon) = \int^{0}_{-\infty} \!\!\!\! d\epsilon_{1} \int^{\infty}_{0} \!\!\!\! d\epsilon_{2} \int^{0}_{-\infty} \!\!\!\! d\epsilon_{3} \,
\dfrac{\hat{\rho}_{L\sigma'}(\epsilon_{1}) \, \hat{\rho}_{L\sigma'}(\epsilon_{2}) \,  
\hat{\rho}_{L'\sigma}(\epsilon_{3})}{(|\epsilon|+|\epsilon_{3}|+|\epsilon_{2}|+|\epsilon_{1}|+|\epsilon_{c}|)^{2}} \, ,
\label{partcleb}
\end{align}
\begin{align}
B_{tLL' \sigma}(\epsilon) = \int^{0}_{-\infty} \!\!\!\! d\epsilon_{1} \int^{\infty}_{0} \!\!\!\! d\epsilon_{2} \int^{0}_{-\infty} \!\!\!\! d\epsilon_{3} \, 
\dfrac{\hat{\rho}_{L -\sigma}(\epsilon_{1}) \, \hat{\rho}_{L' -\sigma}(\epsilon_{2}) \,  
\hat{\rho}_{L'\sigma}(\epsilon_{3})}{(|\epsilon|+|\epsilon_{3}|+|\epsilon_{2}|+|\epsilon_{1}|+|\epsilon_{c}|)^{2}} \, ,
\label{partclebt}
\end{align}
\begin{align}
C_{LL' \sigma'\sigma}(\epsilon) = \int^{0}_{-\infty} \!\!\!\! d\epsilon_{1} \int^{\infty}_{0} \!\!\!\! d\epsilon_{2} \int^{\infty}_{0} \!\!\!\! d\epsilon_{3} \, 
\dfrac{\hat{\rho}_{L\sigma'}(\epsilon_{1}) \, \hat{\rho}_{L\sigma'}(\epsilon_{2}) \,  
\hat{\rho}_{L'\sigma}(\epsilon_{3})}{(|\epsilon_{3}|+|\epsilon|+|\epsilon_{2}|+|\epsilon_{1}|+|\epsilon_{c}|)^{2}} \, ,
\label{holec}
\end{align}
\begin{align}
C_{tLL' \sigma}(\epsilon) = \int^{0}_{-\infty} \!\!\!\! d\epsilon_{1} \int^{\infty}_{0} \!\!\!\! d\epsilon_{2} \int^{\infty}_{0} \!\!\!\! d\epsilon_{3} \, 
\dfrac{\hat{\rho}_{L -\sigma}(\epsilon_{1}) \, \hat{\rho}_{L' -\sigma}(\epsilon_{2}) \,  
\hat{\rho}_{L\sigma}(\epsilon_{3})}{(|\epsilon_{3}|+|\epsilon|+|\epsilon_{2}|+|\epsilon_{1}|+|\epsilon_{c}|)^{2}} \, .
\label{partclect}
\end{align}
$\hat{\rho}_{L\sigma}(\epsilon)$ is the LDOS measured from the Fermi level, and $\epsilon_{c}$ is the correlation energy given by Eq. (\ref{corre}).

As seen from Eq. (\ref{avznum}), the deviation of the MDF from the FDF near the Fermi level  depends on the electronic structure mainly via the LDOS at $\epsilon_{F}$ and the particle and hole functions (\ref{partcleb})-(\ref{partclect}).  These functions are enhanced when the $d$ local densities of states are high near the Fermi level.

It is convenient to consider the MDF projected onto the orbital $L$ in order to understand the role of the $s$, $p$, and $d$ electrons.  The projected MDF (PMDF) are defined by $\langle n_{kL\sigma}\rangle=\sum_{n}\langle n_{kn\sigma}\rangle\vert u_{Ln\sigma}(\bm{k})\vert^{2}$  in which the energy $\epsilon_{kn\sigma}$ at the rhs has been replaced with $\epsilon_{kL\sigma}=\sum_{n}\epsilon_{kn\sigma}\vert u_{Ln\sigma}(\bm{k})\vert^{2}$, $i.e.,$ a common energy band projected onto the orbital $L$.  Here $u_{Ln\sigma}(\bm{k})$ is the eigenvector for a given $\bm{k}$ point.
\begin{equation}
\langle n_{kL\sigma} \rangle=f(\tilde{\epsilon}_{kL\sigma})+\frac{N\langle\tilde{O}_{i}^{\dagger}(\delta{n}_{kL\sigma}) \tilde{O}_{i}\rangle_{0}}{1+\langle{\tilde{O_i}^\dagger\tilde{O_i}}\rangle_0}\,.
\label{nklsigma}
\end{equation}
The PMDF depends on the momentum $\bm{k}$ only via $\tilde{\epsilon}_{kL\sigma}$.  Then the $s$, $p$, and $d$ partial MDF are defined by
\begin{align}
\langle n_{kl\sigma} \rangle = \frac{1}{(2l+1)} \sum_{m} \langle n_{kL\sigma} \rangle . 
\label{partialnk}
\end{align}

We can define the quasiparticle weight $Z_{L\sigma}$ for the electrons  with orbital $L$ by means of the jump of $\langle n_{kL\sigma}\rangle$ at $\bm{k}_{F}$ averaged over the Fermi surface. 
\begin{align}
Z_{L\sigma}=1+\frac{\overline{\delta(N\langle\tilde{O}_{i}^{\dagger} (\delta{n}_{kL\sigma}) \tilde{O}_{i}\rangle_{0})}_{k_{F}}}{1+\langle{\tilde{O_i}^\dagger\tilde{O_i}}\rangle_0}\,.
\label{zlsigma}
\end{align}
Then we find the following sum rule. 
\begin{align}
Z_{\sigma} = \frac{1}{D}\sum_{L}Z_{L\sigma} = \frac{1}{D} \sum_{l=0}^{2} (2l+1) Z_{l\sigma} \,.
\label{sumzl}
\end{align}
Here $Z_{l\sigma}=\sum_{m} Z_{L\sigma} / (2l+1)$ are the partial quasiparticle weights for the $l \, (= s, p, d)$ electrons with spin $\sigma$, which are obtained from $\langle n_{kl\sigma} \rangle$.

The mass enhancement factor (MEF) $(m^{\ast}/m)_{\sigma}$ for $\sigma$-spin electrons is given by
\begin{align}
\left( \frac{m^{\ast}}{m} \right)_{\sigma} = \frac{1}{Z_{\sigma}} \, ,
\label{meffsigma}
\end{align}
and the average MEF $m^{\ast}/m$ is given by
\begin{align}
\frac{m^{\ast}}{m} = \dfrac{\rho_{\uparrow}(\epsilon_{F})}{\rho(\epsilon_{F})} 
\left( \frac{m^{\ast}}{m} \right)_{\uparrow} + \dfrac{\rho_{\downarrow}(\epsilon_{F})}{\rho(\epsilon_{F})} \left( \frac{m^{\ast}}{m} \right)_{\downarrow} \, .
\label{avmeff}
\end{align}
Here $\rho(\epsilon)$ is the total DOS per atom.

\section{Numerical Results for Fe, Co, and Ni}

We applied the spin-polarized MLA to the ferromagnetic transition metals, Fe, Co, and Ni to clarify the quantitative aspects of the theory.

\subsection{Coulomb and exchange energy parameters in the LDA+U}

In the numerical calculations, we adopted simplified intra-orbital Coulomb interaction $U_{0}= \sum_{m} U_{mm}/5$, inter-orbital Coulomb interaction $U_{1}=\sum_{m,m'}^{\prime} U_{mm'}/20$, and the average exchange interaction energy $J=\sum_{m,m'}^{\prime} J_{mm'}/20$.
These Coulomb and exchange energy parameters were obtained in our previous papers as follows~\cite{kake08-2, kake10}, using the averaged parameters $\bar{U}$ and $\bar{J}$ obtained from the LDA+U method.
We assumed first that
\begin{align}
\bar{U} &= \frac{1}{25} \sum_{m,m'} U_{mm'} = \frac{1}{5} U_{0} + \frac{4}{5} U_{1},  
\label{baru1} \\
\bar{J} \,\, &= \frac{1}{20} {\sum_{m,m'}}^{\prime} J_{mm'} \, =  J.
\label{barj1}
\end{align}
Using the sum rule $U_{0}=U_{1}+2J$ for the cubic system as well as the above relations, we find the following expressions.
\begin{align}
U_{0} &= \bar{U} + \frac{8}{5} \, \bar{J} ,  \\
U_{1} &= \bar{U} - \frac{2}{5} \, \bar{J} ,  \\
J \,\, &= \bar{J} ,
\label{ujcase1}
\end{align}
which we call case I hereafter.

However, we note that the relation (\ref{baru1}) is not obious in the LDA+U scheme, and only $\bar{U}$ and $\bar{U}-\bar{J}$ are obtained via the change of the LDA charge density potential~\cite{kake08,anisimov93}.  Thus instead of Eqs. (\ref{baru1}) and (\ref{barj1}), we may use the following relations to determine $U_{0}$, $U_{1}$, and $J$.
\begin{align}
\bar{U} = \alpha U_{0} + (1-\alpha) U_{1},  
\label{baru2} \\
\bar{U} - \bar{J} =  U_{1} - J .  \hspace{11mm}
\label{barj2}
\end{align}
When we adopt Eqs. (\ref{baru2}) and (\ref{barj2}) as well as the sum rule $U_{0}=U_{1}+2J$, we obtain the expressions.
\begin{align}
U_{0} &= \bar{U} + \frac{2(1-\alpha)}{2\alpha+1} \, \bar{J} , 
\label{ujcase211} \\
U_{1} &= \bar{U} - \frac{2\alpha}{2\alpha+1} \, \bar{J} , 
\label{ujcase212} \\
J \,\, &= \frac{1}{2\alpha+1} \, \bar{J} .
\label{ujcase213}
\end{align}
If we adopt a naive value $\alpha=1/5$ ({\it i.e.}, Eq. (\ref{baru1})) in the above expressions, we obtain 
\begin{align}
U_{0} &= \bar{U} + \frac{8}{7} \, \bar{J} ,  \\
U_{1} &= \bar{U} - \frac{2}{7} \, \bar{J} ,  \\
J \,\, &= \frac{5}{7} \, \bar{J} .
\label{ujcase22}
\end{align}
We call the above choice of $U_{0}$, $U_{1}$, and $J$, case II.

In the numerical calculations, we studied the two cases mentioned above using the LDA+U values, $\bar{U}=0.169$ Ry and $\bar{J}=0.066$ Ry for Fe~\cite{anisimov97}, $\bar{U}=0.245$ Ry and $\bar{J}=0.069$ Ry for fcc Co~\cite{bdyo89,mann67}, and $\bar{U}=0.221$ Ry and $\bar{J}=0.066$ for Ni~\cite{anisimov97}.  It should be noted that $J$ for case II are close to those obtained by the cRPA (constraint Random Phase Approximation), e.g., $J=0.047$ Ry (case II) for Fe is compared with $J=0.046$ Ry in the cRPA~\cite{miya08}.

In the self-consistent calculations, we first assume the local charge $\langle n_{iL} \rangle$ ($l=2$) and magnetization $\langle m_{iL} \rangle$ ($l=2$) as well as chemical potential $\epsilon_{F}$.  Using the relations (\ref{ansatzn}) and (\ref{ansatzm}), we calculate the spin-polarized energy bands $\epsilon_{kn\sigma}$ for one-electron Hartree-Fock Hamiltonian with potential (\ref{vare}), so that one can calculate various Hartree-Fock averages in the self-consistent equations.  
Next, we solve iteratively the self-consistent equations (\ref{corre}) for correlation energy and (\ref{eqvp2}) for variational parameters, starting from their lowest order values in Coulomb interactions.  
After solving these equations, we recalculate $\langle n_{iL} \rangle$, $\langle m_{iL} \rangle$, and $\epsilon_{F}$ according to Eqs. (\ref{nil}), (\ref{mil}), and (\ref{nefermi}).  This procedure is repeated until the self-consistency of $\langle n_{iL} \rangle$, $\langle m_{iL} \rangle$, and $\epsilon_{F}$ is achieved.
%
%
\begin{table}[tbh]
\caption{Ground-state spin magnetizations per atom for Fe, fcc Co, and Ni calculated by various methods.  Expt. : experimental data~\cite{danan68,besun70,reck69}, LDA : results for the local density approximation in the density functional theory (DFT)~\cite{moruzzi78,moru86,moru86-2}, GGA: results for the generalized gradient approximation in the DFT~\cite{coco05,wu01}, MLA(I) : MLA results calculated with use of $U_{0}$, $U_{1}$, and $J$ for case I, MLA(II) : MLA results for case II.  The LDA and GGA results are obtained at experimental equilibrium volumes.
\vspace{5mm} }
\label{table-m}
\begin{tabular}{cccccc}
\hline
Element  & Expt. ($\mu_{\rm B}$) & LDA & GGA & MLA(I) & MLA(II) \\ \hline
\vspace*{1mm}
Fe & 2.12 & 2.15 & 2.46 & 2.55 & 2.45  \\
Co &1.69 & 1.56 & 1.66 & 1.71 & 1.74 \\
Ni & 0.57 & 0.59 & 0.66 & 0.54 & 0.59 \\
\hline 
\end{tabular}
\end{table}
%
%
%
%
\begin{figure}[hptb]
\begin{center}
\includegraphics[width=8.5cm]{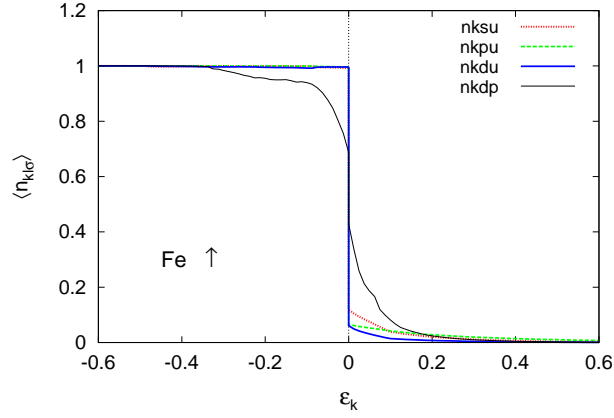}
\end{center}
\vspace{1cm}
\caption{ (Color online) \ Projected momentum distribution functions (PMDF)  
$\langle n_{kl\sigma}\rangle$ for up-spin electrons as a function 
of the energy $\epsilon_{k}$ 
( = $\epsilon_{kL\sigma} - \epsilon_{\rm F}$)  for Fe. 
Dotted curve (red) : the PMDF for up-spin $s$ electrons ($l=0$), 
dashed curve (green) : the PMDF for up-spin $p$ electrons ($l=1$), solid curve (blue) : 
the PMDF for up-spin $d$ electrons ($l=2$), thin solid curve (black) : the PMDF for $d$ electrons in the paramagnetic state.
}
\label{fig-fenklu}
\end{figure}
%
%
%
%
\begin{figure}[hptb]
\begin{center}
\includegraphics[width=8.5cm]{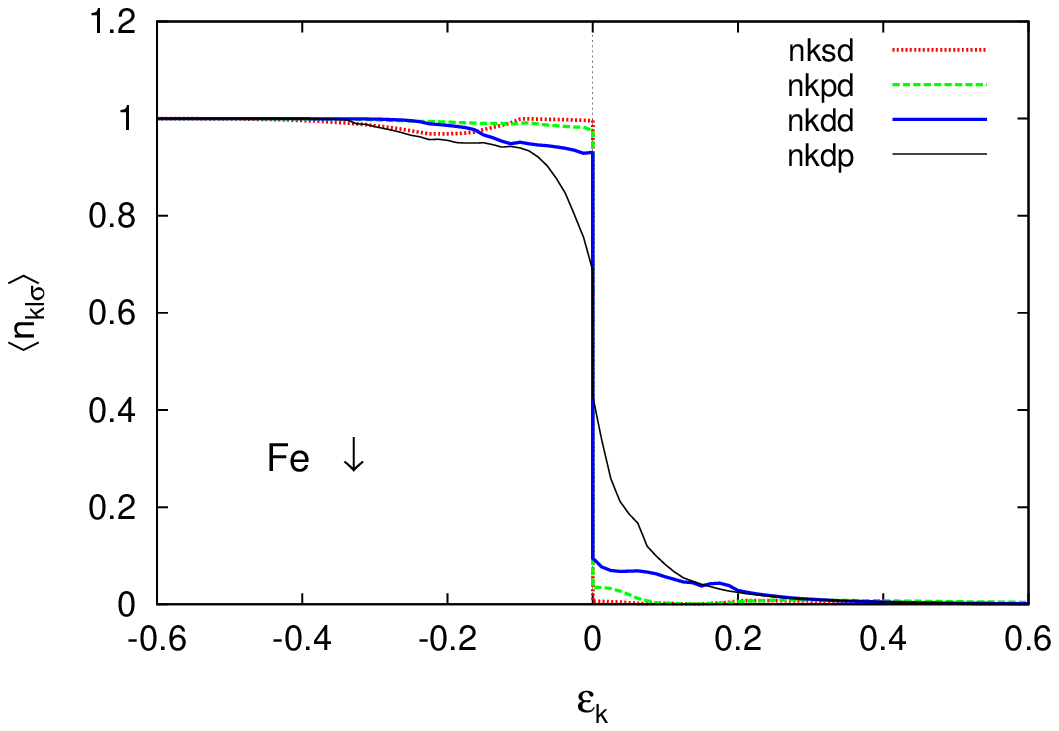}
\end{center}
\vspace{1cm}
\caption{ (Color online) \ 
The PMDF  
$\langle n_{kl\sigma}\rangle$ vs energy $\epsilon_{k}$ curves for down-spin electrons of Fe. 
Dotted curve (red) : the PMDF for down-spin $s$ electrons ($l=0$), 
dashed curve (green) : the PMDF for down-spin $p$ electrons ($l=1$), solid curve (blue) : 
the PMDF for down-spin $d$ electrons ($l=2$), thin solid curve (black) : the PMDF for $d$ electrons in the paramagnetic state.
}
\label{fig-fenkld}
\end{figure}
%
%
%
%
\begin{figure}[hptb]
\begin{center}
\includegraphics[width=9.0cm]{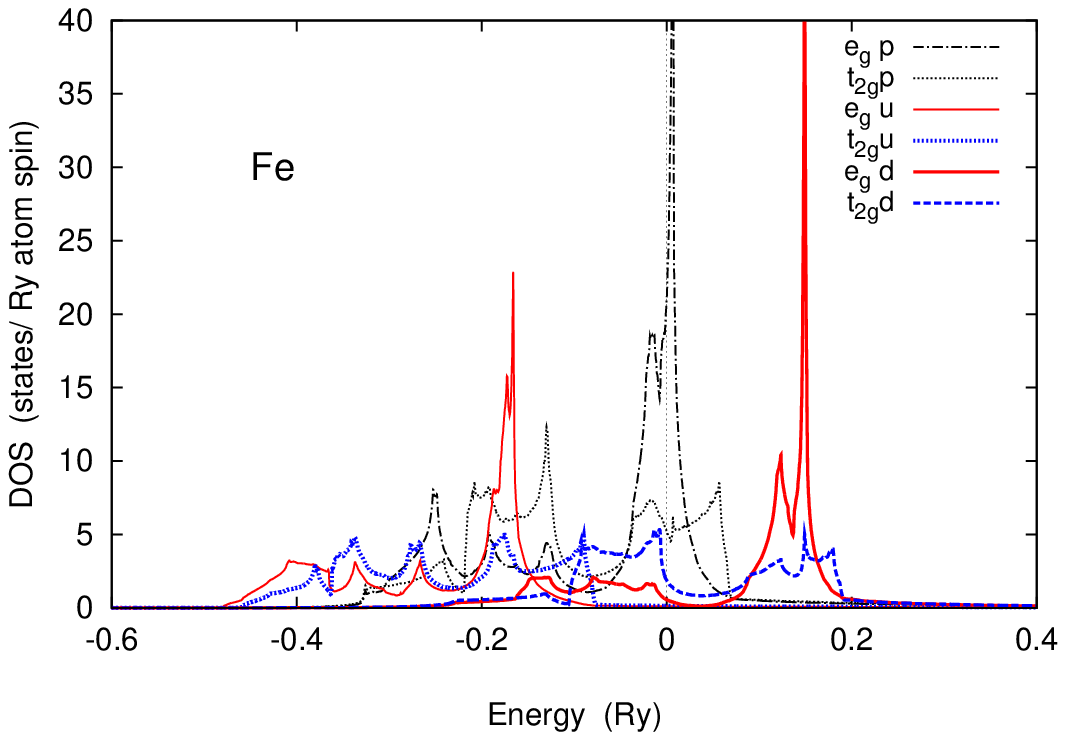}
\end{center}
\vspace{1cm}
\caption{ (Color online) \ 
Densities of states (DOS) for e${}_{\rm g}$ and t${}_{\rm 2g}$ orbitals in the ferro- and para-magnetic Fe.  Note that the energy is measured from the Fermi level.  Thin solid curve (red) : the DOS for e${}_{\rm g}$ up-spin electrons, dotted curve (blue) : the DOS for t${}_{\rm 2g}$ up-spin electrons, solid curve (red) : the DOS for e${}_{\rm g}$ down-spin electrons, dashed curve (blue) : the DOS for t${}_{\rm 2g}$ down-spin electrons, dot-dashed curve (black) : the DOS for e${}_{\rm g}$ electrons in the paramagnetic state, thin dotted curve (black) : the DOS for t${}_{\rm 2g}$ electrons in the paramagnetic state.  These are obtained from the Hartree-Fock one-electron Hamiltonian with use of the self-consistent potential (\ref{vare}) and Coulomb interaction energy parameters for case II. 
}
\label{fig-fedosef}
\end{figure}
%
%

\subsection{Ground-state magnetizations}

Calculated ground-state spin magnetizations are summarized in Table I together with the LDA and GGA results as well as the experimental values.  We obtained 2.55 $\mu_{\rm B}$ (case I) and 2.45 $\mu_{\rm B}$ (case II) for bcc Fe.  Smaller spin magnetization in case II is mainly caused by the reduction of exchange energy, {\it i.e.}, $J=0.066$ Ry (case I)  $\rightarrow 0.046$ Ry (case II).  These values are consistent with the zero-temperature magnetizations 2.58 $\mu_{\rm B}$ (case I) and 2.39 $\mu_{\rm B}$ (approximate case II with $J=0.046$ Ry), which are obtained by an extrapolation of the magnetization-temperature curves calculated with use of the dynamical CPA (Coherent Potential Approximation)~\cite{kake11}.  
The MLA magnetization 2.55 $\mu_{\rm B}$ for case I is, however, overestimated as compared with the experimental value 2.12 $\mu_{\rm B}$.  Calculated value 2.45 $\mu_{\rm B}$ for case II is better and comparable to the GGA results 2.46 $\mu_{\rm B}$.   
In order to reproduce the experimental value 2.12 $\mu_{\rm B}$ using the Coulomb interactions  (\ref{ujcase211})-(\ref{ujcase213}), we have to choose $\alpha \approx 0.6$ instead of $\alpha=0.2$.
%
%
\begin{figure}[htbh]
\begin{center}
\includegraphics[width=8.5cm]{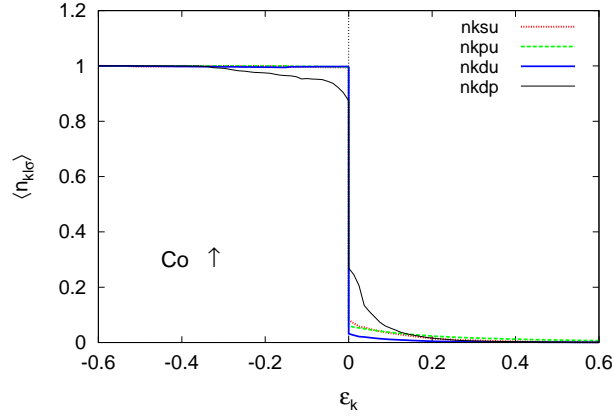}
\end{center}
\vspace{1cm}
\caption{ (Color online) \ 
The PMDF  
$\langle n_{kl\sigma}\rangle$ vs energy $\epsilon_{k}$ curves for up-spin electrons of fcc Co. 
Dotted curve (red) : the PMDF for up-spin $s$ electrons ($l=0$), 
dashed curve (green) : the PMDF for up-spin $p$ electrons ($l=1$), solid curve (blue) : the PMDF for up-spin $d$ electrons ($l=2$), thin solid curve (black) : the PMDF for $d$ electrons in the paramagnetic state.
}
\label{fig-conklu}
\end{figure}
%
%
%
%
\begin{figure}[htbh]
\begin{center}
\includegraphics[width=8.5cm]{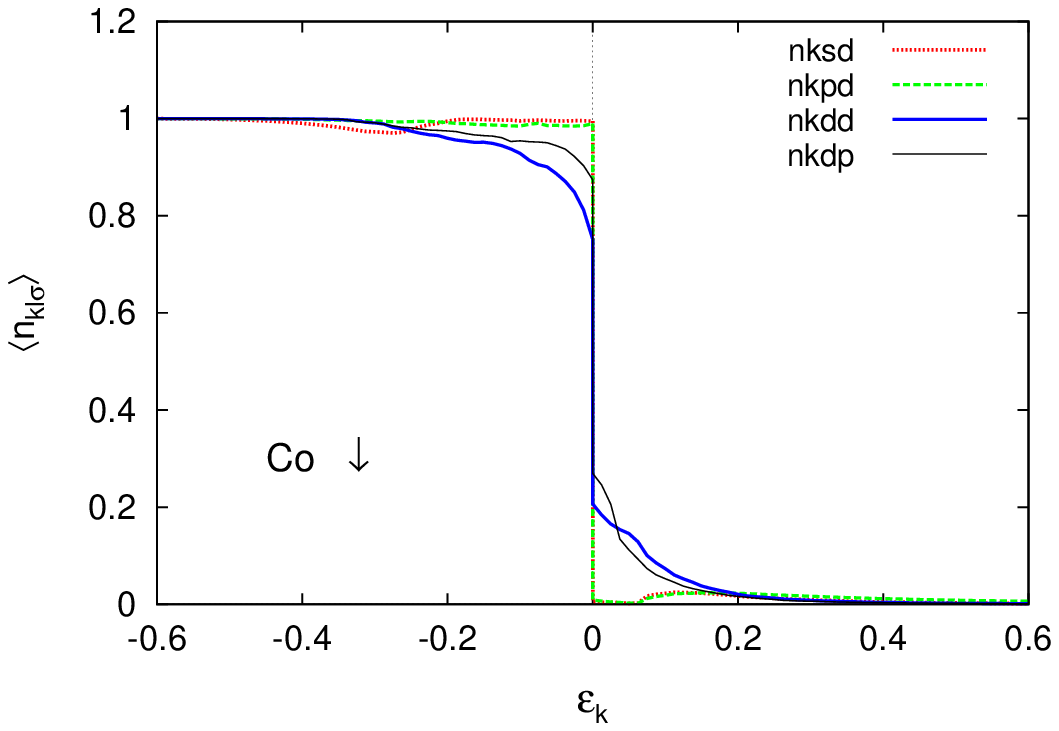}
\end{center}
\vspace{1cm}
\caption{ (Color online) \ 
The PMDF  
$\langle n_{kl\sigma}\rangle$ vs energy $\epsilon_{k}$ curves for down-spin electrons of Co. 
Dotted curve (red) : the PMDF for down-spin $s$ electrons ($l=0$), 
dashed curve (green) : the PMDF for down-spin $p$ electrons ($l=1$), solid curve (blue) : the PMDF for down-spin $d$ electrons ($l=2$), thin solid curve (black) : the PMDF for $d$ electrons in the paramagnetic state.
}
\label{fig-conkld}
\end{figure}
%
%
%
%
\begin{figure}[htbp]
\begin{center}
\includegraphics[width=9.0cm]{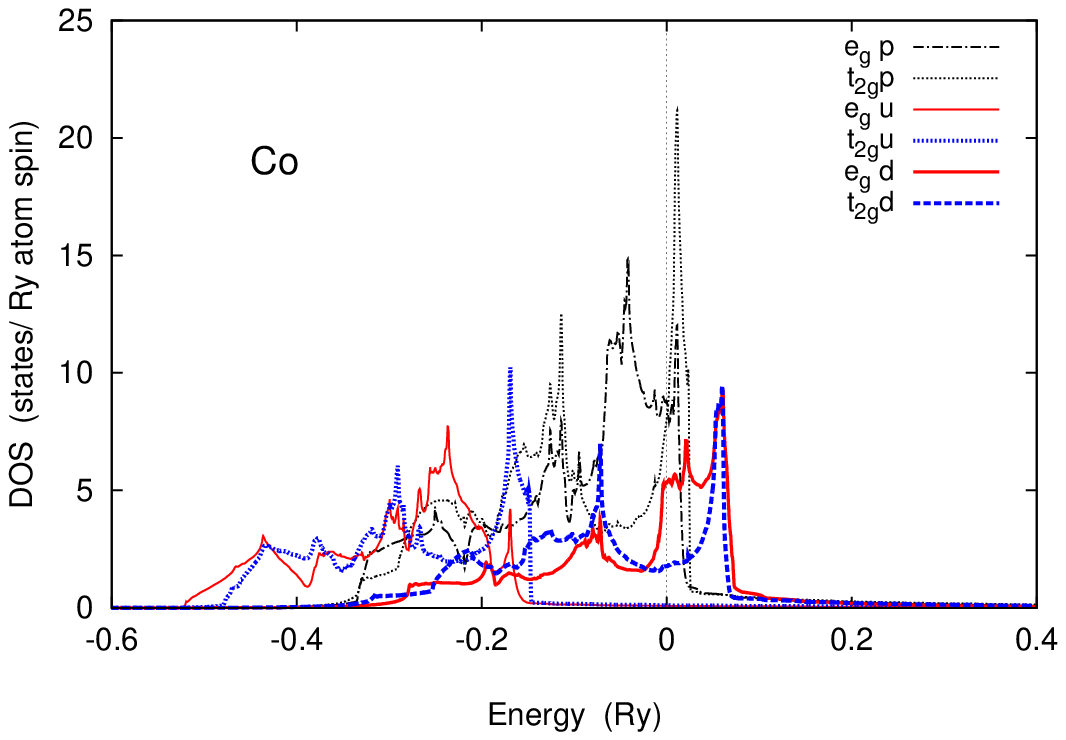}
\end{center}
\vspace{1cm}
\caption{ (Color online) 
The DOS for e${}_{\rm g}$ and t${}_{\rm 2g}$ orbitals in the ferro- and para-magnetic Co.  Thin solid curve (red) : DOS for e${}_{\rm g}$ up-spin electrons, dotted curve (blue) : DOS for t${}_{\rm 2g}$ up-spin electrons, solid curve (red) : DOS for e${}_{\rm g}$ down-spin electrons, dashed curve (blue) : DOS for t${}_{\rm 2g}$ down-spin electrons, dot-dashed curve (black) : DOS for e${}_{\rm g}$ electrons in the paramagnetic state, thin dotted curve (black) : DOS for t${}_{\rm 2g}$ electrons in the paramagnetic state. 
}
\label{fig-codosef}
\end{figure}
%
%

In the case of fcc Co, we obtained the magnetization 1.71 $\mu_{\rm B}$ (case I) and 1.74 $\mu_{\rm B}$ (case II).  
Small enhancement of magnetization in case II as compared with the case I is caused by the detailed balance between the Coulomb interactions and the kinetic energy of electrons.
When we change the Coulomb interactions from the case I ($U_{0}=0.356$, $U_{1}=0.218$, $J=0.069$ Ry) to the case II ($U_{0}=0.324$, $U_{1}=0.226$, $J=0.049$ Ry),  the inter-orbital Coulomb interactions are increased.  Then we find the charge transfer from the $d$  to $sp$ electrons to reduce the Coulomb energy loss.   Accordingly, the polarization of $sp$ electrons being antiparallel to the magnetization is reduced by 0.02 $\mu_{\rm B}$.  Since the reduction of the $d$ electron number is mainly caused by the e${}_{\rm g}$ down-spin electrons on the Fermi level, the e${}_{\rm g}$ polarization is increased by 0.14.  The t${}_{2g}$ electrons slightly increase and their spin polarization decreases by 0.13, so that we find the net increment of magnetization 0.01$\mu_{\rm B}$ for $d$ electrons.  These changes in $sp$ and $d$ polarizations explain the change in magnetization  from 1.71 to 1.74 $\mu_{\rm B}$.
Calculated results are larger than those obtained by the DFT, 1.56 $\mu_{\rm B}$ (LDA) and 1.66 $\mu_{\rm B}$ (GGA), but seem to be better agreement with the experimental value 1.69 $\mu_{\rm B}$. 

We obtained the ground-state magnetizations of Ni, 0.54 $\mu_{\rm B}$ for case I and 0.59 $\mu_{\rm B}$ for case II.  The latter of case II is again larger than the case I.
When we change the Coulomb interactions from case I to case II,  the inter-orbital Coulomb interactions increase and the charge transfer from the $d$ to $sp$ orbitals occurs as discussed in fcc Co.
The change of $sp$ spin polarization is negligible in case of Ni. 
Since the e${}_{\rm g}$ electrons are almost occupied below the Fermi level, the reduction of $d$ electrons is realized by the t${}_{\rm 2g}$ down-spin electrons.
Then, the same amount of magnetization is increased by 0.05 $\mu_{\rm B}$, 
This explains the increment of magnetization for case II. Present results 0.54 (case I) and 0.59 $\mu_{\rm B}$ (case II) are smaller than the GGA value 0.66 $\mu_{\rm B}$, but seem to be better agreement with the experimental value 0.57 $\mu_{\rm B}$.
%
%
\begin{figure}[htbh]
\begin{center}
\includegraphics[width=8.5cm]{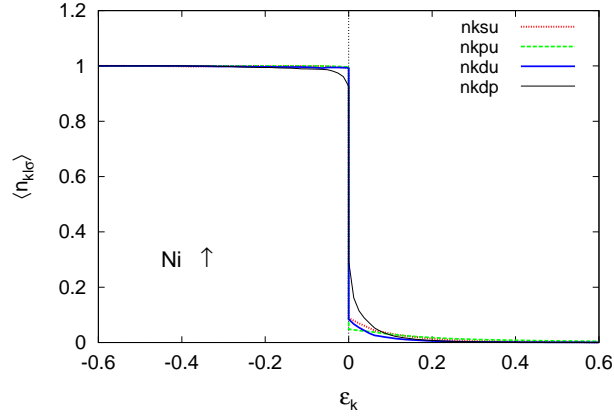}
\end{center}
\vspace{1cm}
\caption{ (Color online)  \ 
The PMDF $\langle n_{kl\sigma}\rangle$ vs energy $\epsilon_{k}$ curves for up-spin electrons of Ni. 
Dotted curve (red) : the PMDF for up-spin $s$ electrons ($l=0$), 
dashed curve (green) : the PMDF for up-spin $p$ electrons ($l=1$), solid curve (blue) : the PMDF for up-spin $d$ electrons ($l=2$), thin solid curve (black) : the PMDF for $d$ electrons in the paramagnetic state.
}
\label{fig-ninklu}
\end{figure}
%
%
%
%
\begin{figure}[htbh]
\begin{center}
\includegraphics[width=8.5cm]{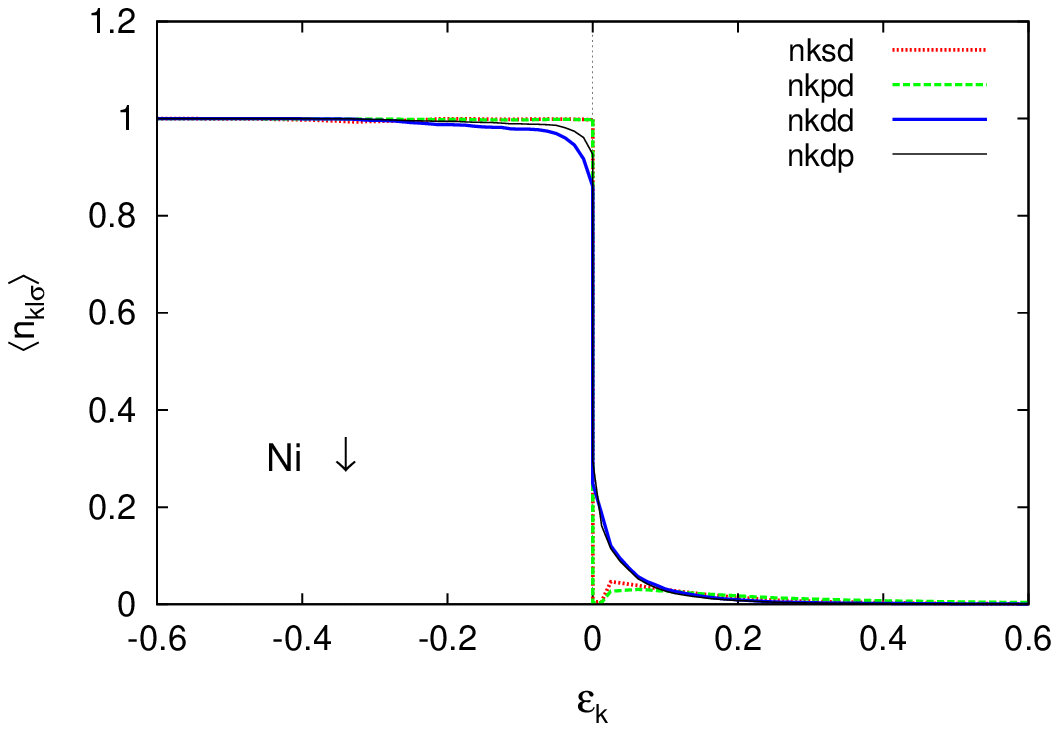}
\end{center}
\vspace{1cm}
\caption{ (Color online)  \ 
The PMDF  
$\langle n_{kl\sigma}\rangle$ vs energy $\epsilon_{k}$ curves for down-spin electrons of Ni. 
Dotted curve (red) : PMDF for down-spin $s$ electrons ($l=0$), 
dashed curve (green) : PMDF for down-spin $p$ electrons ($l=1$), solid curve (blue) : PMDF 
for down-spin $d$ electrons ($l=2$), thin solid curve (black) : PMDF for $d$ electrons in the paramagnetic state. 
}
\label{fig-ninkld}
\end{figure}
%
%
%
%
\begin{figure}[htbp]
\begin{center}
\includegraphics[width=9.0cm]{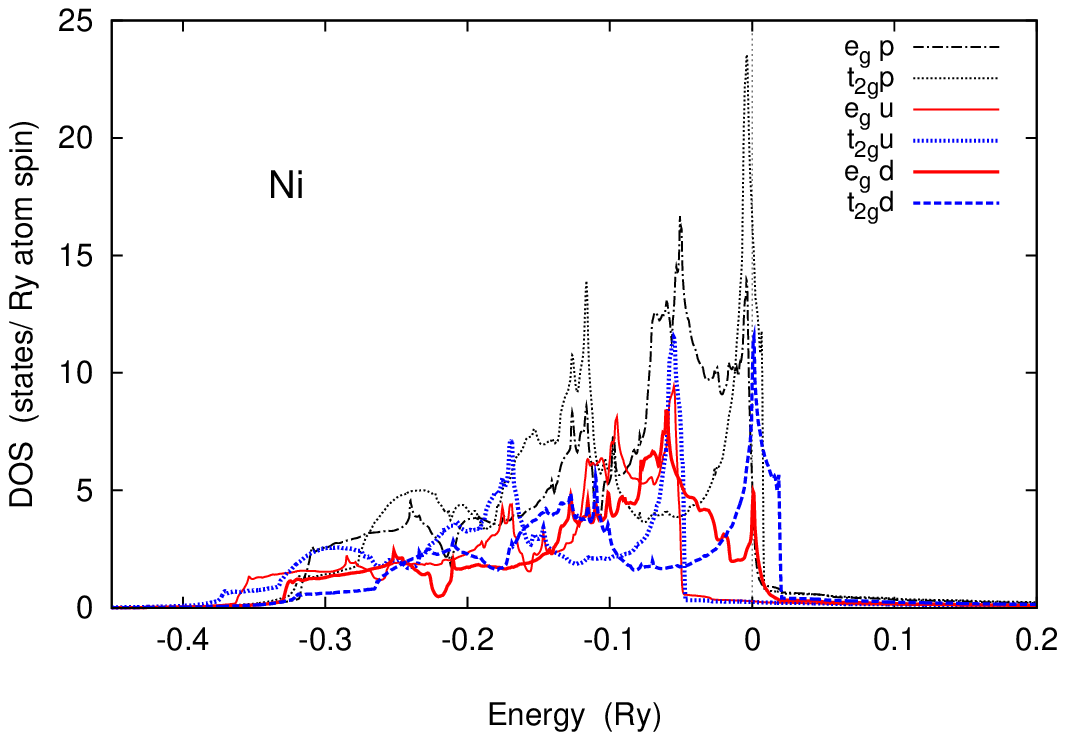}
\end{center}
\vspace{1cm}
\caption{ (Color online) \ 
The DOS for e${}_{\rm g}$ and t${}_{\rm 2g}$ orbitals in the ferro- and para-magnetic Ni.  Thin solid curve (red) : DOS for e${}_{\rm g}$ up-spin electrons, dotted curve (blue) : DOS for t${}_{\rm 2g}$ up-spin electrons, solid curve (red) : DOS for e${}_{\rm g}$ down-spin electrons, dashed curve (blue) : DOS for t${}_{\rm 2g}$ down-spin electrons, dot-dashed curve (black) : DOS for e${}_{\rm g}$ electrons in the paramagnetic state, thin dotted curve (black) : DOS for t${}_{\rm 2g}$ electrons in the paramagnetic state. 
}
\label{fig-nidosef}
\end{figure}
%
%

\subsection{Momentum distribution functions}

The momentum distribution functions (MDF) of transition metals in the paramagnetic state have been investigated in details in our previous papers~\cite{kake17}.  In the paramagnetic Fe, we found a significant deviation of the MDF from the Fermi distribution function (FDF) because of the flat e${}_{\rm g}$ bands on the Fermi level. Here we discuss the ferromagnetic case.  We present in Figs. 1 and 2 the calculated projected MDF (PMDF) for up- and down-spin electrons in the ferromagnetic Fe (case II).  In the ferromagnetic state, the deviations of the MDF from the FDF are strongly suppressed for both the up- and down-spin $d$ electrons as compared with the paramagnetic ones.   In order to see the reason from the viewpoint of electronic structure, we present in Fig. 3 the orbital resolved densities of states (DOS) obtained in the self-consistent calculations for case II.  As seen in Fig. 3, the e${}_{\rm g}$ band on the Fermi level in the paramagnetic state splits into the up and down bands in the ferromagnetic state because of the  exchange splitting.  It reduces much the densities of states on the Fermi level, and thus reduces the deviation of the PMDF for $d$ electrons.  On the other hand, the MDF projected onto the $sp$ orbitals for up-spin states are considerably enhanced above $\epsilon_{F}$ in the ferromagnetic state.  In the ferromagnetic state, the e${}_{\rm g}$ and t${}_{\rm 2g}$ up-spin bands sink far below the Fermi level as shown in Fig. 3, while the $sp$ electron states on the Fermi level considerably increase, so that the deviation of the PMDF for up-spin sp electrons is enhanced via the hybridization with $d$ electrons and becomes comparable to that for the $d$ electrons. 

Calculated PMDF for the fcc ferromagnetic Co are presented in Figs. 4 and 5. 
We find again the strong suppression of the deviation from the FDF for the up-spin $d$ electrons in Fig. 4, because the e${}_{\rm g}$ and t${}_{\rm 2g}$ up-spin bands are located far below the Fermi level (see Fig. 6).  However, the deviation of the PMDF for the down-spin $d$ electrons remains unchanged as seen in Fig. 5.
Note that the Fermi level of Co is increased as compared with that of Fe because the conduction electron number is increased by one, so that the Fermi level is located on the high DOS of e${}_{\rm g}$ down-spin electrons as seen in Fig. 6, and thus the PMDF for the down-spin $d$ electrons does not much change even in the ferromagnetic state.

The same behavior of the PMDF is seen in the ferromagnetic Ni as shown in Figs. 7 and 8, though the deviations from the FDF are limited near the Fermi level.  We find the e${}_{\rm g}$ and t${}_{\rm 2g}$ up-spin bands far below the Fermi level in the ferromagnetic Ni as shown in Fig. 9.  It leads to a small deviation from the FDF for up-spin $d$ electrons.  On the other hand, the peaks in the e${}_{\rm g}$ and t${}_{\rm 2g}$ down-spin DOS are still located on the Fermi level. This leads to the deviation of the PMDF for down-spin $d$ electrons, which is comparable to the paramagnetic one.

\subsection{Mass enhancement factors}

We calculated the mass enhancement factors (MEF) from the PMDF at the Fermi level.  The results are presented in Table II.
In Fe, the MEF for both the up- and down-spin electrons in the ferromagnetic state are reduced because of the disappearance of the e${}_{\rm g}$ flat bands on the Fermi level due to exchange splitting.  We obtained the average MEF $m^{\ast}/m = 1.12$ for case I and $1.11$  for case II in the ferromagnetic Fe.  We also calculated the MEF using the Coulomb and exchange energy parameters (\ref{ujcase211})-(\ref{ujcase213}) with $\alpha=0.6$ leading to a reasonable magnetization 2.14 $\mu_{\rm B}$, but found that the average MEF hardly changes as compared with the case II; $m^{\ast}/m = 1.11$ for $\alpha=0.6$.
%
%
\begin{table}[htbh]
\caption{Calculated mass enhancement factors (MEF) for Fe, fcc Co, and Ni.
$(m^{\ast}/m)_{\rm p}$: MEF in the paramagnetic state, $(m^{\ast}/m)_{\sigma}$: MEF for electrons with spin $\sigma \, ( \, = \uparrow, \downarrow)$, $m^{\ast}/m$: average MEF in the ferromagnetic state.  I and II in the second column imply the results for case I and case II, respectively.
\vspace{5mm} }
\label{table-meff}
\begin{tabular}{ccccc}
\hline
Element  & $(m^{\ast}/m)_{\rm p}$ & $(m^{\ast}/m)_{\uparrow}$ & $(m^{\ast}/m)_{\downarrow}$ & $m^{\ast}/m$  \\ \hline
Fe \, I & 1.71 & 1.08 & 1.13 & 1.12  \\
\hspace*{6mm} II & 1.44 & 1.11 & 1.11 & 1.11  \\
Co \, I & 1.29 & 1.05 & 1.35 & 1.33  \\
\hspace*{6mm} II & 1.21 & 1.06 & 1.27 & 1.24  \\ 
Ni \, I & 1.26 & 1.09 & 1.28 & 1.27  \\
\hspace*{6mm} II & 1.23 & 1.09 & 1.24 & 1.23  \\ 
\hline 
\end{tabular}
\end{table}
%
%
%
%
\begin{figure}[tbp]
\begin{center}
\includegraphics[width=9.0cm]{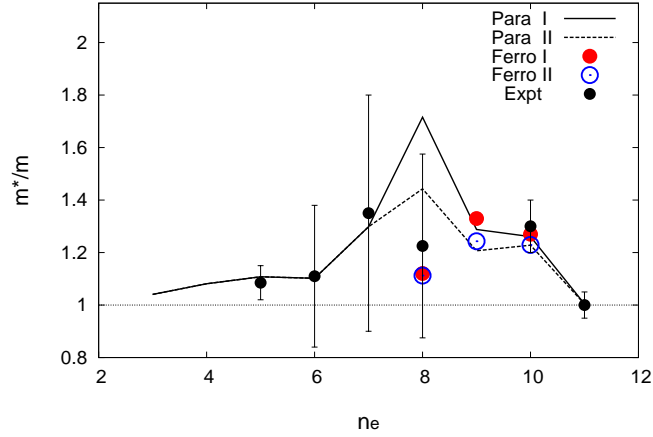}
\end{center}
\vspace{1cm}
\caption{(Color online) \ Calculated mass enhancement factors $m^{\ast}/m$ of transition metals as a function of conduction electron number $n_{\rm e}$.  Solid line: the paramagnetic results~\cite{kake17} for case I, dashed line: the paramagnetic results for case II, closed circles (red) : the ferromagnetic results of $m^{\ast}/m$ for case I, open circles (blue) : the ferromagnetic results of $m^{\ast}/m$ for case II.  Small closed-circles (black) with error bars are the experimental results~\cite{papacon15,knapp72,beck70,cheng60,zimm61}.  Note that the electron-phonon contributions have been subtracted in the data by using the high-temperature experimental data or theoretical results~\cite{weiss58, pepp01, mcmil68, grin69, jarl02}.
}
\label{fig-mecmp}
\end{figure}
%
%

In the case of fcc Co, the MEF of up-spin electrons is reduced since the up-spin $d$ bands sink below the Fermi level, while the MEF for down-spin electrons is rather enhanced by 0.06 as compared with the paramagnetic one because the e${}_{\rm g}$ DOS for down-spin electrons still remain on the Fermi level in the ferromagnetic state (see Fig. 6).  The average MEF  in the ferromagnetic Co turns out $m^{\ast}/m = 1.33$ ($1.24$) in case I (case II), which is comparable to $(m^{\ast}/m)_{\rm p}=1.29$ ($1.21$).

The MEF in the ferromagnetic Ni decreases for up-spin electrons, but hardly changes for down-spin electrons because of the same reason as in fcc Co, though it is caused by the t${}_{\rm 2g}$ down electrons in the case of ferromagnetic Ni.  Finally, we find $m^{\ast}/m = 1.27$ ($1.23$) in case I (case II), which is compared with $(m^{\ast}/m)_{\rm p}=1.26$ ($1.23$) in the paramagnetic state.

Calculated MEF in the ferromagnetic state as well as those in the paramagnetic state are summarized in Fig. 10.  Experimental results are obtained from the specific heat data~\cite{papacon15,knapp72,beck70,cheng60,zimm61} by subtracting the contributions of the electron-phonon interaction using the high-temperature experimental data or theoretical results~\cite{weiss58, pepp01, mcmil68, grin69, jarl02}.  Ferromagnetic results for Fe, Co, and Ni seem to be consistent with the experimental data, though the latter have considerable ambiguity due to uncertainty in subtraction of the electron-phonon contributions.

\section{Summary}

We have extended the momentum-dependent local ansatz approach (MLA) to the spin polarized case.  The MLA describes electron correlations at zero temperature from the weak to the intermediate Coulomb interaction regime quantitatively, and allows us to make realistic calculations of various physical quantities at zero temperature on the same footing.

In order to clarify the quantitative aspects of the spin-polarized MLA, we performed the self-consistent calculations of the ground-state spin magnetizations and the momentum distribution functions (MDF) for Fe, fcc Co, and Ni, using two sets of Coulomb ($U_{0}$, $U_{1}$) and exchange ($J$) interaction energy parameters: case I with $J=\bar{J}$ and case II with $J=5\bar{J}/7$, where $\bar{J}$ is the exchange energy parameters in the LDA+U method.

We obtained the  magnetizations 2.55 (2.45) $\mu_{\rm B}$ for Fe, 1.71 (1.74) $\mu_{\rm B}$  for Co, and 0.54 (0.59) $\mu_{\rm B}$ for Ni in case I (II), respectively.  These results are comparable to the results for the GGA in the DFT ({\it i.e.}, 2.46 $\mu_{\rm B}$ for Fe, 1.66 $\mu_{\rm B}$ for Co, and 0.66 $\mu_{\rm B}$ for Ni), though the calculated spin magnetization of Fe is considerably larger than the experimental value 2.12 $\mu_{\rm B}$.
Since calculated magnetization for Fe is considerably sensitive for the choice of Coulomb interaction parameters, we need more quantitative method to determine the Coulomb interactions for the present Hamiltonian.  It is also left for future investigations to perform more detailed self-consistent calculations including the optimization of the potential parameters in the Hartree-Fock wavefunction in order to clarify the accuracy of the MLA for the ground-state magnetization in Fe.

We calculated the MDF to clarify the spin polarization effects in Fe, Co, and Ni.  In the ferromagnetic Fe, the correlation effects on the MDF are suppressed for both spin electrons because the e${}_{\rm g}$ flat band on the Fermi level is split into the up and down bands away from the Fermi level.  Thus the deviation of the MDF from the FDF is much reduced for both spin electrons, and the mass enhancement factor (MEF) decreases from 1.71 (1.44) to 1.13 (1.11) in case I (II).   

In Co and Ni, correlation effects on the MDF for up-spin electrons are suppressed for the same reason as in the ferromagnetic Fe.  But those for down-spin electrons remain unchanged and even increase in the ferromagnetic state, because the Fermi level is located on the e${}_{\rm g}$ and t${}_{\rm 2g}$ down-spin bands due to larger conduction electron number as compared with that of Fe.  Consequently, we obtain basically the same MEF as those in the paramagnetic state: 1.33 (1.24) in Co and 1.27 (1.23) in Ni for case I (II).  Calculated MEF in the ferromagnetic state explain the experimental results obtained from the specific heat data.  

Although the present results of MEF are consistent with the experimental data, there is considerable ambiguity in subtraction of the electron-phonon contribution from the raw data.
In order to examine the quantitative aspects of the low energy excitations more precisely, it might be better to calculate the quasiparticle energy rather than the quasiparticle weight because one can compare the results with the angle-resolved photoemission spectroscopy data (ARPES) in more details.  The extension of the MLA to the excited states is possible using the methods developed in the wavefunction methods~\cite{hors83,bune03}.  Calculations of the excited states with use of the MLA along this line are in progress.

\clearpage

\appendix

\section{Matrix Elements in the Self-Consistent Equations (\ref{eqvp2})}

 In this Appendix, we present the matrix elements $\tilde{Q}_{\tau LL' \sigma\sigma'}^{(\alpha\alpha')}$, $P_{\tau LL' \sigma\sigma'}^{(\alpha'\alpha)}$, and $K_{\tau LL' \sigma\sigma'}^{(\alpha)}$ in the self-consistent equations (\ref{eqvp2}) for variational parameters.
First, $\tilde{Q}_{\tau LL' \sigma\sigma'}^{(\alpha\alpha')}$ are given as
\begin{align}
\tilde{Q}^{(00)}_{LL \downarrow\uparrow} \ \ &= \tilde{Q}_{LL \downarrow\uparrow} \, ,
\hspace{40mm}
\tilde{Q}^{(11)}_{LL' \sigma\sigma'} = \tilde{Q}_{LL' \sigma\sigma'} ,   \\ 
\tilde{Q}^{(12)}_{l LL' \sigma\sigma'} &= \tilde{Q}^{(21)}_{l LL' \sigma\sigma'} = -\frac{1}{4}\sigma\sigma' \tilde{Q}_{LL' \sigma\sigma'} ,
\hspace{7mm}
\tilde{Q}^{(22)}_{l LL' \sigma\sigma'} = \frac{1}{16} \, \tilde{Q}_{LL' \sigma\sigma'} \, ,   \\
\tilde{Q}^{(22)}_{t LL' \sigma-\sigma} &= \frac{1}{4} \, \tilde{Q}_{tLL' \sigma} \, .
\label{tq}
\end{align}
The other matrix elements $(\alpha\alpha')$ of $\tilde{Q}_{\tau LL' \sigma\sigma'}^{(\alpha\alpha')}$ vanish, and 
\begin{align}
\tilde{Q}_{LL' \sigma\sigma'} &= -\int_{0}^{\infty} dt \, dt' e^{i\epsilon_{c} (t+t')} \Bigl[ a_{L'\sigma'}(-t-t') \, b_{L'\sigma'}(t+t') \, a_{L\sigma}(-t-t') \, b_{L1\sigma}(t+t') 
\nonumber \\
&\hspace{3.3cm} 
- a_{L'\sigma'}(-t-t') \, b_{L'\sigma'}(t+t') \, a_{L1\sigma}(-t-t') \, b_{L\sigma}(t+t')  
\nonumber \\
&\hspace{3.3cm} 
+ a_{L'\sigma'}(-t-t') \, b_{L'1\sigma'}(t+t') \, a_{L\sigma}(-t-t') \, b_{L\sigma}(t+t')  
\nonumber \\
&\hspace{3.3cm} 
- a_{L'1\sigma'}(-t-t') \, b_{L'\sigma'}(t+t') \, a_{L\sigma}(-t-t') \, b_{L\sigma}(t+t') 
\Bigr]  \nonumber \\
&+ \epsilon_{c} 
\int_{0}^{\infty} dt \, dt' e^{i\epsilon_{c} (t+t')} 
\, a_{L'\sigma'}(-t-t') \, b_{L'\sigma'}(t+t')  \, a_{L\sigma}(-t-t') \, b_{L\sigma}(t+t') \, ,  
\label{tqqs}
\end{align}
\begin{align}
\tilde{Q}_{tLL' \sigma} &= -\int_{0}^{\infty} dt \, dt' e^{i\epsilon_{c} (t+t')} \Bigl[ a_{L'-\sigma}(-t-t') \, b_{L'\sigma}(t+t') \, a_{L\sigma}(-t-t') \, b_{L1-\sigma}(t+t') 
\nonumber \\
&\hspace{3.3cm} 
- a_{L'-\sigma}(-t-t') \, b_{L'\sigma}(t+t') \, a_{L1\sigma}(-t-t') \, b_{L-\sigma}(t+t')  
\nonumber \\
&\hspace{3.3cm} 
+ a_{L'-\sigma}(-t-t') \, b_{L'1\sigma}(t+t') \, a_{L\sigma}(-t-t') \, b_{L-\sigma}(t+t')  
\nonumber \\
&\hspace{3.3cm} 
- a_{L'1-\sigma}(-t-t') \, b_{L'\sigma}(t+t') \, a_{L\sigma}(-t-t') \, b_{L-\sigma}(t+t') 
\Bigr]  \nonumber \\
&+ \epsilon_{c} 
\int_{0}^{\infty} dt \, dt' e^{i\epsilon_{c} (t+t')} 
\, a_{L'-\sigma}(-t-t') \, b_{L'\sigma}(t+t')  \, a_{L\sigma}(-t-t') \, b_{L-\sigma}(t+t') \,.
\label{tqtqs}
\end{align}
Here
\begin{align}
a_{L\sigma}(t)=\int_{-\infty}^{\infty}{d\epsilon \ {e^{-i\epsilon t}}f(\tilde{\epsilon})\ {\rho}_{L\sigma}(\epsilon)}\,,
\label{alsigma}
\end{align}
\begin{align}
b_{L\sigma}(t)=\int_{-\infty}^{\infty}{d\epsilon \ {e^{-i\epsilon t}}f(-\tilde{\epsilon})\ {\rho}_{L\sigma}(\epsilon)}\,,
\label{blsigma}
\end{align}
\begin{align}
a_{1L\sigma}(t)=\int_{-\infty}^{\infty}{d\epsilon \ {e^{-i\epsilon t}}f(\tilde{\epsilon})\ \epsilon \ {\rho}_{L\sigma}(\epsilon)}\,,
\label{a1lsigma}
\end{align}
\begin{align}
b_{1L\sigma}(t)=\int_{-\infty}^{\infty}{d\epsilon\ {e^{-i\epsilon t}} f(-\tilde{\epsilon}) \ \epsilon \ {\rho}_{L\sigma}(\epsilon)}\,.
\label{b1lsigma}
\end{align}
$f(\tilde{\epsilon})$ is the Fermi distribution function and 
$\tilde{\epsilon}$ denotes the energy measured from the Fermi level.  $\rho_{L\sigma}(\epsilon)$ at the rhs is the local density of states for orbital $L$ and spin $\sigma$.
\begin{align}
\rho_{L\sigma}(\epsilon)=\sum_{kn}\vert\langle iL\vert kn\rangle_{\sigma} \vert^{2}\ \delta{(\epsilon-\epsilon_{kn\sigma})}\,.
\label{doslsigma}
\end{align}

Next, the matrix elements $P_{\tau LL' \sigma\sigma'}^{(\alpha'\alpha)}$ in Eq. (\ref{eqvp2}) are given as
\begin{align}
P^{(00)}_{LL \downarrow\uparrow} \ \ &= P_{LL \downarrow\uparrow} \, ,
\hspace{40mm}
P^{(11)}_{LL' \sigma\sigma'} = P_{LL' \sigma\sigma'} ,   \\ 
P^{(12)}_{l LL' \sigma\sigma'} &= -P^{(21)}_{l LL' \sigma\sigma'} = -\frac{1}{4}\sigma\sigma' P_{LL' \sigma\sigma'} ,
\hspace{5mm}
P^{(22)}_{l LL' \sigma\sigma'} = -\frac{1}{16} \, P_{LL' \sigma\sigma'} \, ,   \\
P^{(22)}_{t LL' \sigma-\sigma} &= -\frac{1}{4} \, P_{tLL' \sigma} \, .
\label{phio}
\end{align}
The other elements $(\alpha\alpha')$ of $P_{\tau LL' \sigma\sigma'}^{(\alpha'\alpha)}$ vanish, and 
\begin{align}
P_{LL' \sigma\sigma'} = i \int_{0}^{\infty} dt e^{i\epsilon_{c} t} 
\, a_{L\sigma}(-t) \, a_{L'\sigma'}(-t) \, b_{L\sigma}(t) \, b_{L'\sigma'}(t) \, ,   \nonumber \\
P_{tLL' \sigma} = i \int_{0}^{\infty} dt e^{i\epsilon_{c} t} 
\, a_{L\sigma}(-t) \, a_{L'-\sigma}(-t) \, b_{L-\sigma}(t) \, b_{L'\sigma}(t) \, .
\label{ppthio}
\end{align}

The matrix elements $K_{\tau LL' \sigma\sigma'}^{(\alpha)}$ are given as follows.

\begin{align}
K_{LL\downarrow\uparrow}^{(0)} &= {{U}_{LL}^{(0)}}^{2} \Omega_{LL\downarrow\uparrow} \, \tilde{\lambda}_{0LL}   \nonumber \\
&+ \sum_{L' (\neq L)} \, {U}_{LL'}^{(1)} \, \sum_{\sigma\sigma'} \, \big( U_{LL'}^{(1)} - \frac{1}{4}\sigma\sigma' U^{(2)}_{LL'}  \big) \, M_{LL'\sigma\sigma'} \, \tilde{\lambda}^{(\sigma\sigma')}_{1LL'}  \nonumber \\
& - \frac{1}{4} \sum_{L' (\neq L)} \, U^{(2)}_{LL'} \, \sum_{\sigma\sigma'} \, \sigma\sigma' \big( U^{(1)}_{LL'} - \frac{1}{4} \sigma\sigma' {U}^{(2)}_{LL'} \, \big) M_{LL'\sigma\sigma'} \, \tilde{\lambda}^{(\sigma\sigma')}_{2lLL'}   \nonumber \\
&+ \frac{1}{4} \sum_{L' (\neq L)} {{U}^{(2)}_{LL'}}^{2} \sum_{\sigma} \hat{M}_{tLL'\sigma} \, 
\tilde{\lambda}^{(\sigma -\sigma)}_{2tLL'} \, ,
\label{k0du}
\end{align}
\begin{align}
K^{(1)}_{LL'\sigma\sigma'} &= \Big( {U}^{(1)}_{LL'} - \frac{1}{4} \, \sigma\sigma' \, U^{(2)}_{LL'} \Big) \, 
 \big( U^{(0)}_{LL} M_{LL'\sigma\sigma'} \, \tilde{\lambda}_{0LL} + U^{(0)}_{L'L'} M_{L'L\sigma'\sigma} \, \tilde{\lambda}_{0L'L'}  \big)  \nonumber \\
&+ U^{(1)}_{LL'} \, \big( 
U^{(0)}_{LL} \, \Xi_{L'LL\sigma'\sigma -\sigma} \, \tilde{\lambda}^{(-\sigma\sigma')}_{1LL'}
+ U^{(0)}_{L'L'} \, \Xi_{LL'L'\sigma\sigma' -\sigma'} \, \tilde{\lambda}^{(\sigma -\sigma')}_{1LL'}
\big) \nonumber \\
&+ \sum_{L'' (\neq L,L')} U^{(1)}_{LL''} \sum_{\sigma''} \Big( 
U^{(1)}_{L'L''} + \frac{1}{4}\sigma'\sigma'' U^{(2)}_{L'L''} \Big) \, 
\Xi_{LL'L''\sigma\sigma'\sigma''} \, \tilde{\lambda}^{(\sigma\sigma'')}_{1LL''}  \nonumber \\
&+ \sum_{L'' (\neq L,L')} U^{(1)}_{L'L''} \sum_{\sigma''} \Big( 
U^{(1)}_{LL''} + \frac{1}{4}\sigma\sigma'' U^{(2)}_{LL''} \Big) \, 
\Xi_{L'LL''\sigma'\sigma\sigma''} \, \tilde{\lambda}^{(\sigma'\sigma'')}_{1L'L''} \nonumber \\
&+ U^{(1)}_{LL'} \, \Big( 
U^{(1)}_{LL'} + \frac{1}{4} \, \sigma\sigma' U^{(2)}_{LL'} \Big) \, 
\Omega_{LL'\sigma\sigma'} \, \tilde{\lambda}^{(\sigma\sigma')}_{1LL'} \nonumber \\
&+ \frac{1}{4} \, \sigma\sigma' \, U^{(2)}_{LL'} \, \big( 
U^{(0)}_{LL} \, \Xi_{L'LL\sigma'\sigma -\sigma} \, \tilde{\lambda}^{(-\sigma\sigma')}_{2lLL'}
+ U^{(0)}_{L'L'} \, \Xi_{LL'L'\sigma\sigma' -\sigma'} \, \tilde{\lambda}^{(\sigma -\sigma')}_{2lLL'} \, \big) \nonumber \\
&- \frac{1}{4} \, \sigma\sigma' \sum_{L'' (\neq L, L')} U^{(2)}_{L'L''} \sum_{\sigma''} \, \big(
\sigma'\sigma'' U^{(1)}_{L'L''} + \frac{1}{4} U^{(2)}_{L'L''}  \big) \, 
\Xi_{LL'L''\sigma\sigma' \sigma''} \, \tilde{\lambda}^{(\sigma \sigma'')}_{2lLL''}    
\nonumber \\
&- \frac{1}{4} \, \sigma\sigma' \sum_{L'' (\neq L, L')} U^{(2)}_{L'L''} \sum_{\sigma''} \big(
\sigma\sigma'' U^{(1)}_{LL''} + \frac{1}{4} U^{(2)}_{LL''}  \big) \, 
\Xi_{L'LL''\sigma'\sigma\sigma''} \, \tilde{\lambda}^{(\sigma' \sigma'')}_{2lL'L''} 
\nonumber \\
&- \frac{1}{4} \, U^{(2)}_{LL'} \, \big(
\sigma\sigma' U^{(1)}_{LL'} + \frac{1}{4} U^{(2)}_{LL'}  \big) \, 
\Omega_{LL'\sigma\sigma'} \, \tilde{\lambda}^{(\sigma \sigma')}_{2lLL'} 
\nonumber \\
&+ \frac{1}{8} \, {U^{(2)}_{LL'}}^{2} \, \big( 
\hat{Z}_{1LL'\sigma} \tilde{\lambda}^{(\sigma -\sigma)}_{2tLL'} + 
\hat{Z}_{2LL'\sigma} \tilde{\lambda}^{(-\sigma \sigma)}_{2tLL'} -
\hat{Z}_{3LL'\sigma} \tilde{\lambda}^{(-\sigma \sigma)}_{2tLL'} -
\hat{Z}_{4LL'\sigma} \tilde{\lambda}^{(\sigma -\sigma)}_{2tLL'} \big) 
\nonumber \\
&+ \frac{1}{8} \, \sigma\sigma' {U^{(2)}_{LL'}}^{2} \, \big( 
\hat{Z}_{1LL'\sigma} \tilde{\lambda}^{(\sigma -\sigma)}_{2tLL'} + 
\hat{Z}_{2LL'\sigma} \tilde{\lambda}^{(-\sigma \sigma)}_{2tLL'} +
\hat{Z}_{3LL'\sigma} \tilde{\lambda}^{(-\sigma \sigma)}_{2tLL'} +
\hat{Z}_{4LL'\sigma} \tilde{\lambda}^{(\sigma -\sigma)}_{2tLL'} \big) \, ,
\label{k1ll}
\end{align}
\begin{align}
K^{(2)}_{lLL'\sigma\sigma'} &=  - \frac{1}{4} \sigma\sigma' K^{(1)}_{LL'\sigma\sigma'} \, ,
\label{k2ll}
\end{align}
\begin{align}
K^{(2)}_{tLL'\sigma\sigma'} &=  
\frac{1}{4} \, U^{(2)}_{LL'} \, \big( \, 
U^{(0)}_{LL} \hat{M}_{tLL'\sigma} \, \tilde{\lambda}_{0LL} + 
U^{(0)}_{L'L'} \hat{M}_{tL'L-\sigma} \, \tilde{\lambda}_{0L'L'}  \big)  \nonumber \\
&+ \frac{1}{4} \, U^{(1)}_{LL'} \, U^{(2)}_{LL'} \, \big( \, 
\hat{Z}_{1LL'\sigma} \tilde{\lambda}^{(\sigma \sigma)}_{1LL'} + 
\hat{Z}_{2LL' -\sigma} \tilde{\lambda}^{(-\sigma -\sigma)}_{1LL'} -
\hat{Z}_{3LL' -\sigma} \tilde{\lambda}^{(-\sigma \sigma)}_{1LL'} -
\hat{Z}_{4LL'\sigma} \tilde{\lambda}^{(\sigma -\sigma)}_{1LL'} \big) \nonumber \\
&- \frac{1}{16} \, {U^{(2)}_{LL'}}^{2} \, \big( \, 
\hat{Z}_{1LL'\sigma} \tilde{\lambda}^{(\sigma \sigma)}_{2lLL'} + 
\hat{Z}_{2LL' -\sigma} \tilde{\lambda}^{(-\sigma -\sigma)}_{2lLL'} +
\hat{Z}_{3LL' -\sigma} \tilde{\lambda}^{(-\sigma \sigma)}_{2lLL'} +
\hat{Z}_{4LL'\sigma} \tilde{\lambda}^{(\sigma -\sigma)}_{2lLL'} \big) \nonumber \\
&- \frac{1}{4} \, U^{(1)}_{LL'} U^{(2)}_{LL'} \, \big( \, 
\tilde{Z}_{12LL'\sigma} - \tilde{Z}_{34LL'\sigma} \, \big) \, \tilde{\lambda}^{(\sigma -\sigma)}_{2tLL'}
\nonumber \\
&+ \frac{1}{8} \sum_{L'' (\neq L, L')} U^{(2)}_{LL''} \, U^{(2)}_{L'L''} \, \big( \, 
\tilde{\Xi}_{LL'L''\sigma} \, \tilde{\lambda}^{(\sigma -\sigma)}_{2tLL''} + 
\tilde{\Xi}_{L'LL'' -\sigma} \, \tilde{\lambda}^{(-\sigma \sigma)}_{2tL'L''} \, \big)
\nonumber \\
&- \frac{1}{16} \, {U^{(2)}_{LL'}}^{2} \, \big( \, 
\tilde{Z}_{12LL'\sigma} + \tilde{Z}_{34LL'\sigma} \, \big) \, \tilde{\lambda}^{(\sigma -\sigma)}_{2tLL'}
+ \frac{1}{8} \, {U^{(2)}_{LL'}}^{2} \, \tilde{Z}^{\prime}_{34LL'\sigma} \, 
\tilde{\lambda}^{(-\sigma \sigma)}_{2tLL'} \, .
\label{k2tll}
\end{align}
Here 
\begin{align}
\Omega_{LL'\sigma\sigma'} &=
\int_{0}^{\infty} \! \! \! \! dt dt' e^{i\epsilon_{c} (t+t')} \Bigl[ 
a_{L'\sigma'}(-t) \, b_{L'\sigma'}(t+t') \, a_{L\sigma}(-t-t') \, b_{L\sigma}(t) \, 
a_{L'\sigma'}(-t') \, b_{L\sigma}(t')   \nonumber \\
&\hspace{2.5cm} +
a_{L'\sigma'}(-t-t') \, b_{L'\sigma'}(t) \, a_{L\sigma}(-t) \, b_{L\sigma}(t+t') \, 
b_{L'\sigma'}(t') \, a_{L\sigma}(-t')   \nonumber \\
&\hspace{2.5cm} -
a_{L'\sigma'}(-t) \, b_{L'\sigma'}(t+t') \, a_{L\sigma}(-t) \, b_{L\sigma}(t+t') \, 
a_{L'\sigma'}(-t') \, a_{L\sigma}(-t')   \nonumber \\
&\hspace{2.5cm} -
a_{L'\sigma'}(-t-t') \, b_{L'\sigma'}(t) \, a_{L\sigma}(-t-t') \, b_{L\sigma}(t) \, 
b_{L'\sigma'}(t') \, b_{L\sigma}(t') \, \Bigl] \, ,
\label{omgll}
\end{align}
\begin{align}
\Xi_{LL'L''\sigma\sigma'\sigma''} &=
- \int_{0}^{\infty} \! \! \! \! dt dt' e^{i\epsilon_{c} (t+t')} 
a_{L'\sigma'}(-t) \, b_{L'\sigma'}(t) \, a_{L\sigma}(-t-t') \, b_{L\sigma}(t+t') \, 
a_{L''\sigma''}(-t') \, b_{L''\sigma''}(t')  \, ,
\label{xilll}
\end{align}
\begin{align}
M_{LL'\sigma\sigma'} &= \Xi_{LLL'\sigma-\sigma\sigma'}  \, ,
\label{mll}
\end{align}
\begin{align}
\hat{M}_{tLL'\sigma} &=
- \int_{0}^{\infty} \! \! \! \! dt dt' e^{i\epsilon_{c} (t+t')} 
a_{L -\sigma}(-t) \, b_{L -\sigma}(t+t') \, a_{L\sigma}(-t-t') \, b_{L\sigma}(t) \, 
a_{L' -\sigma}(-t') \, b_{L'\sigma}(t')  \, ,
\label{mhtll}
\end{align}
\begin{align}
\hat{Z}_{1LL'\sigma} &=
- \int_{0}^{\infty} \! \! \! \! dt dt' e^{i\epsilon_{c} (t+t')} 
a_{L'\sigma}(-t) \, b_{L'\sigma}(t+t') \, a_{L\sigma}(-t-t') \, b_{L\sigma}(t) \, 
a_{L' -\sigma}(-t') \, b_{L -\sigma}(t')  \, ,
\label{zh1ll}
\end{align}
\begin{align}
\hat{Z}_{2LL'\sigma} &=
- \int_{0}^{\infty} \! \! \! \! dt dt' e^{i\epsilon_{c} (t+t')} 
a_{L'\sigma}(-t-t') \, b_{L'\sigma}(t) \, a_{L\sigma}(-t) \, b_{L\sigma}(t+t') \, 
b_{L' -\sigma}(t') \, a_{L -\sigma}(-t')  \, ,
\label{zh2ll}
\end{align}
\begin{align}
\hat{Z}_{3LL'\sigma} \ =
- \int_{0}^{\infty} \! \! \! \! dt dt' e^{i\epsilon_{c} (t+t')} 
a_{L' -\sigma}(-t) \, b_{L' -\sigma}(t+t') \, a_{L\sigma}(-t) \, b_{L\sigma}(t+t') \, 
a_{L'\sigma}(-t') \, a_{L -\sigma}(-t')  \, ,  \hspace{5mm}
\label{zh3ll}
\end{align}
\begin{align}
\hat{Z}_{4LL'\sigma} &=
- \int_{0}^{\infty} \! \! \! \! dt dt' e^{i\epsilon_{c} (t+t')} 
a_{L' -\sigma}(-t-t') \, b_{L' -\sigma}(t) \, a_{L\sigma}(-t-t') \, b_{L\sigma}(t) \, 
b_{L'\sigma}(t') \, b_{L -\sigma}(t')  \, ,
\label{zh4ll}
\end{align}
\begin{align}
\tilde{\Xi}_{LL'L''\sigma} &=
- \int_{0}^{\infty} \! \! \! \! dt dt' e^{i\epsilon_{c} (t+t')} 
a_{L' -\sigma}(-t) \, b_{L'\sigma}(t) \, a_{L\sigma}(-t-t') \, b_{L -\sigma}(t+t') \, 
a_{L'' -\sigma}(-t') \, b_{L''\sigma}(t')  \, ,     \ \ \ 
\label{xitlll}
\end{align}
\begin{align}
\tilde{Z}_{12LL'\sigma} &=
- \int_{0}^{\infty} \! \! \! \! dt dt' e^{i\epsilon_{c} (t+t')} \, \Big[
a_{L' -\sigma}(-t) \, b_{L'\sigma}(t+t') \, a_{L\sigma}(-t-t') \, b_{L -\sigma}(t) \, 
a_{L' -\sigma}(-t') \, b_{L -\sigma}(t')     \,  \nonumber  \\ 
 & \hspace*{25mm} + 
a_{L' -\sigma}(-t-t') \, b_{L'\sigma}(t) \, a_{L\sigma}(-t) \, b_{L -\sigma}(t+t') \, 
a_{L\sigma}(-t') \, b_{L'\sigma}(t')  \,  \Big] \, ,
\label{zt12ll}
\end{align}
\begin{align}
\tilde{Z}_{34LL'\sigma} &=
- \int_{0}^{\infty} \! \! \! \! dt dt' e^{i\epsilon_{c} (t+t')} \, \Big[ 
a_{L' -\sigma}(-t) \, b_{L'\sigma}(t+t') \, a_{L\sigma}(-t) \, b_{L -\sigma}(t+t') \, 
a_{L' -\sigma}(-t') \, a_{L\sigma}(-t')     \,  \nonumber  \\
 & \hspace*{25mm} + 
a_{L' -\sigma}(-t-t') \, b_{L'\sigma}(t) \, a_{L\sigma}(-t-t') \, b_{L -\sigma}(t) \, 
b_{L'\sigma}(t') \, b_{L -\sigma}(t')  \,  \Big] \, ,
\label{zt34ll}
\end{align}
\begin{align}
\tilde{Z}^{\prime}_{34LL'\sigma} &=
- \int_{0}^{\infty} \! \! \! \! dt dt' e^{i\epsilon_{c} (t+t')} 
a_{L' -\sigma}(-t) \, b_{L'\sigma}(t+t') \, a_{L\sigma}(-t) \, b_{L -\sigma}(t+t') \, 
a_{L'\sigma}(-t') \, a_{L -\sigma}(-t')    \,  \nonumber  \\
 & \hspace*{25mm} + 
a_{L' -\sigma}(-t-t') \, b_{L'\sigma}(t) \, a_{L\sigma}(-t-t') \, b_{L -\sigma}(t) \, 
b_{L' -\sigma}(t') \, b_{L\sigma}(t') \,  \Big]  \, .
\label{ztp34ll}
\end{align}

\section{Correlation Corrections to Electron Number and Magnetic Moment}

 In this Appendix B, we present the expressions of $\langle\tilde{O}_{i}^{\dagger}(\delta{n}_{iL})\tilde{O}_{i}\rangle_{0}$, $\langle\tilde{O}_{i}^{\dagger} (\delta{m}_{iL}) \tilde{O}_{i}\rangle_{0}$, and $\langle{\tilde{O_i}^\dagger\tilde{O_i}}\rangle_0$ in Eqs. (\ref{nil}) and (\ref{mil}).  We note that there is no correlation correction to the {\it sp} electrons in the present model.  For {\it d} electrons ($l=2$), we obtain
\begin{align}
\langle \tilde{O}_{i}^{\dagger} (\delta{n}_{iL}) \tilde{O}_{i} \rangle_{0} &=
{{U}_{LL}^{(0)}}^{2}  \, \tilde{\lambda}_{0LL} \, A^{(00)}_{LLL\downarrow\uparrow} 
\, \tilde{\lambda}_{0LL}  
+ \sum_{L' (\neq L)} \, {{U}_{LL'}^{(1)}}^{2} \, \sum_{\sigma\sigma'} 
\, \tilde{\lambda}^{(\sigma\sigma')}_{1L'L} \, A^{(11)}_{L'LL\sigma\sigma'}  
\, \tilde{\lambda}^{(\sigma\sigma')}_{1L'L}  \nonumber \\
& + \, 2 \sum_{L' (\neq L)} \, U^{(1)}_{LL'} \, U^{(2)}_{LL'} \, \sum_{\sigma\sigma'} 
\, \tilde{\lambda}^{(\sigma\sigma')}_{1L'L} \, A^{(12)}_{lL'LL\sigma\sigma'} 
\, \tilde{\lambda}^{(\sigma\sigma')}_{2lL'L}  \nonumber \\
&\hspace{-10mm} + \sum_{L' (\neq L)} \, {U^{(2)}_{LL'}}^{2} \, \sum_{\sigma\sigma'} 
\, \tilde{\lambda}^{(\sigma\sigma')}_{2lL'L} \, A^{(22)}_{lL'LL\sigma\sigma'}
\, \tilde{\lambda}^{(\sigma\sigma')}_{2lL'L} 
+ \sum_{L' (\neq L)} {{U}^{(2)}_{LL'}}^{2} \sum_{\sigma} 
\tilde{\lambda}^{(\sigma -\sigma)}_{2tL'L} A^{(22)}_{tL'LL\sigma -\sigma} \, 
\tilde{\lambda}^{(\sigma -\sigma)}_{2tL'L} \, ,
\label{onilo}
\end{align}
\begin{align}
\langle \tilde{O}_{i}^{\dagger} (\delta{m}_{iL}) \tilde{O}_{i} \rangle_{0} &=
{{U}_{LL}^{(0)}}^{2} \, \tilde{\lambda}_{0LL} \, \hat{A}^{(00)}_{LLL\downarrow\uparrow} 
\, \tilde{\lambda}_{0LL}  
+ \sum_{L' (\neq L)} \, {{U}_{LL'}^{(1)}}^{2} \, \sum_{\sigma\sigma'} 
\, \tilde{\lambda}^{(\sigma\sigma')}_{1L'L} \, \hat{A}^{(11)}_{L'LL\sigma\sigma'}  
\, \tilde{\lambda}^{(\sigma\sigma')}_{1L'L}  \nonumber \\
& + \, 2 \sum_{L' (\neq L)} \, U^{(1)}_{LL'} \, U^{(2)}_{LL'} \, \sum_{\sigma\sigma'} 
\, \tilde{\lambda}^{(\sigma\sigma')}_{1L'L} \, \hat{A}^{(12)}_{lL'LL\sigma\sigma'} 
\, \tilde{\lambda}^{(\sigma\sigma')}_{2lL'L}  \nonumber \\
&\hspace{-10mm} + \sum_{L' (\neq L)} \, {U^{(2)}_{LL'}}^{2} \, \sum_{\sigma\sigma'} 
\, \tilde{\lambda}^{(\sigma\sigma')}_{2lL'L} \, \hat{A}^{(22)}_{lL'LL\sigma\sigma'}  
\, \tilde{\lambda}^{(\sigma\sigma')}_{2lL'L} 
+ \sum_{L' (\neq L)} {{U}^{(2)}_{LL'}}^{2} \sum_{\sigma} 
\, \tilde{\lambda}^{(\sigma -\sigma)}_{2tL'L} \, \hat{A}^{(22)}_{tL'LL\sigma -\sigma} \, 
\tilde{\lambda}^{(\sigma -\sigma)}_{2tL'L} \, .
\label{omilo}
\end{align}
Here
\begin{align}
A^{(00)}_{LLL\downarrow\uparrow} \ \ &= 
A_{LL\downarrow\uparrow}  + A_{LL\uparrow\downarrow} , \, \hspace{15mm}
\hat{A}^{(00)}_{LLL\downarrow\uparrow} = 
A_{LL\downarrow\uparrow} - A_{LL\uparrow\downarrow} \, ,  \nonumber \\
A^{(11)}_{L'LL\sigma\sigma'} &= A_{L'L\sigma\sigma'} ,   \hspace{25mm}
\hat{A}^{(11)}_{L'LL\sigma\sigma'} = \sigma' \, A_{L'L\sigma\sigma'} \, ,  \nonumber \\
A^{(12)}_{lL'LL\sigma\sigma'} &= - \frac{1}{4} \sigma\sigma' A_{L'L\sigma\sigma'} ,  \hspace{12mm}
\hat{A}^{(12)}_{lL'LL\sigma\sigma'} = - \frac{1}{4} \sigma \, A_{L'L\sigma\sigma'} \, ,  \nonumber \\
A^{(22)}_{lL'LL\sigma\sigma'} &= \frac{1}{16} \, A_{L'L\sigma\sigma'} , \hspace{19mm}
\hat{A}^{(22)}_{lL'LL\sigma\sigma'} = \frac{1}{16} \sigma' \, A_{L'L\sigma\sigma'} \, ,  \nonumber \\
A^{(22)}_{tL'LL\sigma -\sigma} &= \frac{1}{4} \, A^{(-)}_{tL'L\sigma} , \hspace{23mm}
\hat{A}^{(22)}_{tL'LL\sigma -\sigma} = \frac{1}{4} \sigma \, A^{(+)}_{tL'L\sigma} \, , 
\label{alllch}
\end{align}
and
\begin{align}
A_{LL'\sigma\sigma'} &=
- \int_{0}^{\infty} \! \! \! \! dt dt' e^{i\epsilon_{c} (t+t')} \Big[
a_{L\sigma}(-t-t') \, b_{L\sigma}(t+t') \, a_{L'\sigma'}(-t-t') \, b_{L'\sigma'}(t) \, 
b_{L'\sigma'}(t')    \nonumber \\  
& \hspace{29mm} - a_{L\sigma}(-t-t') \, b_{L\sigma}(t+t') \, a_{L'\sigma'}(-t) \, b_{L'\sigma'}(t+t') \, a_{L'\sigma'}(-t') 
\Big] \, , 
\label{allss}
\end{align}
\begin{align}
A^{(\pm)}_{tLL'\sigma} &=
- \int_{0}^{\infty} \! \! \! \! dt dt' e^{i\epsilon_{c} (t+t')} \Big[
a_{L\sigma}(-t-t') \, b_{L -\sigma}(t+t') \, a_{L' -\sigma}(-t-t') \, b_{L'\sigma}(t) \, 
b_{L'\sigma}(t')    \nonumber \\  
& \hspace{29mm} \pm a_{L\sigma}(-t-t') \, b_{L -\sigma}(t+t') \, a_{L' -\sigma}(-t) \, b_{L'\sigma}(t+t') \, a_{L' -\sigma}(-t') 
\Big] \, . 
\label{atllss}
\end{align}

The average $\langle \tilde{O}_{i}^{\dagger} \, \tilde{O}_{i} \rangle_{0}$ is given by
\begin{align}
\langle \tilde{O}_{i}^{\dagger} \, \tilde{O}_{i} \rangle_{0} &=
\sum_{L} {{U}_{LL}^{(0)}}^{2}  \, \tilde{\lambda}_{0LL} \, S_{LL\downarrow\uparrow} 
\, \tilde{\lambda}_{0LL}  
+ \sum_{<L L' >} \, {{U}_{LL'}^{(1)}}^{2} \, \sum_{\sigma\sigma'} 
\, \tilde{\lambda}^{(\sigma\sigma')}_{1LL'} \, S_{LL'\sigma\sigma'}  
\, \tilde{\lambda}^{(\sigma\sigma')}_{1LL'}  \nonumber \\
& - \, \frac{1}{2} \sum_{<L L'>} \, U^{(1)}_{LL'} \, U^{(2)}_{LL'} \, \sum_{\sigma\sigma'} 
\sigma\sigma' \, \tilde{\lambda}^{(\sigma\sigma')}_{1LL'} \, S_{LL'\sigma\sigma'} 
\, \tilde{\lambda}^{(\sigma\sigma')}_{2lLL'}  \nonumber \\
&\hspace{-10mm} + \frac{1}{16} \, \sum_{<L L'>} \, {U^{(2)}_{LL'}}^{2} \, \sum_{\sigma\sigma'} 
\, \tilde{\lambda}^{(\sigma\sigma')}_{2lLL'} \, S_{LL'\sigma\sigma'}
\, \tilde{\lambda}^{(\sigma\sigma')}_{2lLL'} 
+ \frac{1}{4} \, \sum_{<L L'>} {{U}^{(2)}_{LL'}}^{2} \sum_{\sigma} 
\tilde{\lambda}^{(\sigma -\sigma)}_{2tLL'} S_{tLL'\sigma} \, 
\tilde{\lambda}^{(\sigma -\sigma)}_{2tLL'} \, .
\label{oob}
\end{align}
Here
\begin{align}
S_{LL'\sigma\sigma'} &=
- \int_{0}^{\infty} \! \! \! \! dt dt' e^{i\epsilon_{c} (t+t')} 
a_{L'\sigma'}(-t-t') \, b_{L'\sigma'}(t+t') \, a_{L\sigma}(-t-t') \, b_{L\sigma}(t+t') \, , 
\label{sllss}
\end{align}
\begin{align}
S_{tLL'\sigma} &=
- \int_{0}^{\infty} \! \! \! \! dt dt' e^{i\epsilon_{c} (t+t')} 
a_{L' -\sigma}(-t-t') \, b_{L'\sigma}(t+t') \, a_{L\sigma}(-t-t') \, b_{L -\sigma}(t+t') \, .
\label{stllss}
\end{align}
The functions $a_{L\sigma}(t)$ and $b_{L\sigma}(t)$ in Eqs. (\ref{allss}), (\ref{atllss}), (\ref{sllss}), and (\ref{stllss}) have been given in Eqs. (\ref{alsigma}) and (\ref{blsigma}).

\section{Expression of $N\langle\tilde{O}_{i}^{\dagger} (\delta{n}_{kn\sigma}) \tilde{O}_{i}\rangle_{0}$
}

The numerator $N\langle \tilde{O}_{i}^{\dagger} (\delta{n}_{kn\sigma}) \tilde{O}_{i}\rangle_{0}$ in the momentum distribution function (\ref{nknsigma}) is given as follows.
\begin{align}
N\langle\tilde{O}_{i}^{\dagger} (\delta{n}_{kn\sigma}) \tilde{O}_{i}\rangle_{0} &=
\sum_{L} {{U}_{LL}^{(0)}}^{2}  \, \tilde{\lambda}_{0LL}^{2} \, |u_{Ln\sigma}(\boldsymbol{k})|^{2} \big(
B_{LL-\sigma\sigma}(\epsilon_{kn\sigma}) f(-\tilde{\epsilon}_{kn\sigma}) - 
C_{LL-\sigma\sigma}(\epsilon_{kn\sigma}) f(\tilde{\epsilon}_{kn\sigma})  \big) 
\nonumber \\ 
&+ \sum_{<L L' >} \,  \sum_{\sigma'} \, \Big[ \big( \sigma U_{LL'}^{(1)}  
\, \tilde{\lambda}^{(\sigma'\sigma)}_{1LL'} - \frac{1}{4} \, \sigma' U_{LL'}^{(2)} \tilde{\lambda}^{(\sigma'\sigma)}_{2lLL'} \, \big)^{2}  \nonumber \\  
& \hspace{22mm} \times \, |u_{L'n\sigma}(\boldsymbol{k})|^{2} \big(
B_{LL'\sigma'\sigma}(\epsilon_{kn\sigma}) f(-\tilde{\epsilon}_{kn\sigma}) - 
C_{LL'\sigma'\sigma}(\epsilon_{kn\sigma}) f(\tilde{\epsilon}_{kn\sigma})  \big)  \nonumber \\
& \hspace{17mm} + \big( \sigma U_{LL'}^{(1)}  
\, \tilde{\lambda}^{(\sigma\sigma')}_{1LL'} - \frac{1}{4} \, \sigma' U_{LL'}^{(2)} \tilde{\lambda}^{(\sigma\sigma')}_{2lLL'} \, \big)^{2}  \nonumber \\  
& \hspace{22mm} \times \, |u_{Ln\sigma}(\boldsymbol{k})|^{2} \big(
B_{L'L\sigma'\sigma}(\epsilon_{kn\sigma}) f(-\tilde{\epsilon}_{kn\sigma}) - 
C_{L'L\sigma'\sigma}(\epsilon_{kn\sigma}) f(\tilde{\epsilon}_{kn\sigma})  \big)
\Big]  \nonumber \\
& \hspace{-20mm} + \frac{1}{4} \, \sum_{<L L'>} \, {U^{(2)}_{LL'}}^{2} \, \Big[
\, \tilde{\lambda}^{(-\sigma\sigma) 2}_{2tLL'} \, \big(
|u_{Ln\sigma}(\boldsymbol{k})|^{2} B_{tLL'\sigma}(\epsilon_{kn\sigma}) f(-\tilde{\epsilon}_{kn\sigma}) - 
|u_{L'n\sigma}(\boldsymbol{k})|^{2} C_{tLL'\sigma}(\epsilon_{kn\sigma}) f(\tilde{\epsilon}_{kn\sigma})  \big) 
\, \nonumber \\
& \hspace{5mm} + \tilde{\lambda}^{(\sigma-\sigma) 2}_{2tLL'} \, \big(
|u_{L'n\sigma}(\boldsymbol{k})|^{2} B_{tL'L\sigma}(\epsilon_{kn\sigma}) f(-\tilde{\epsilon}_{kn\sigma}) - 
|u_{Ln\sigma}(\boldsymbol{k})|^{2} C_{tL'L\sigma}(\epsilon_{kn\sigma}) f(\tilde{\epsilon}_{kn\sigma})  \big)
\Big] \, .
\label{nonkno}
\end{align}
Here $\{ u_{Ln\sigma}(\boldsymbol{k}) \}$ are the eigen vectors for one-electron energy eigen value $\epsilon_{kn\sigma}$, and
\begin{align}
B_{LL'\sigma'\sigma}(\epsilon_{kn\sigma}) &=
- \int_{0}^{\infty} \! \! \! \! dt dt' e^{i (\epsilon_{c}-\epsilon_{kn\sigma}) (t+t')} 
a_{L\sigma'}(-t-t') \, b_{L\sigma'}(t+t') \, a_{L'\sigma}(-t-t') \, , 
\label{bllss}
\end{align}
\begin{align}
C_{LL'\sigma'\sigma}(\epsilon_{kn\sigma}) &=
- \int_{0}^{\infty} \! \! \! \! dt dt' e^{i (\epsilon_{c}+\epsilon_{kn\sigma}) (t+t')} 
a_{L\sigma'}(-t-t') \, b_{L\sigma'}(t+t') \, b_{L'\sigma}(t+t') \, ,   \ \ 
\label{cllss}
\end{align}
\begin{align}
B_{tLL'\sigma}(\epsilon_{kn\sigma}) &=
- \int_{0}^{\infty} \! \! \! \! dt dt' e^{i (\epsilon_{c}-\epsilon_{kn\sigma}) (t+t')}  
a_{L-\sigma}(-t-t') \, b_{L'-\sigma}(t+t') \, a_{L'\sigma}(-t-t') \, , 
\label{btlls}
\end{align}
\begin{align}
C_{tLL'\sigma}(\epsilon_{kn\sigma}) &=
- \int_{0}^{\infty} \! \! \! \! dt dt' e^{i (\epsilon_{c}+\epsilon_{kn\sigma}) (t+t')} 
a_{L-\sigma}(-t-t') \, b_{L'-\sigma}(t+t') \, b_{L\sigma}(t+t') \, .   \ \ 
\label{ctlls}
\end{align}

\end{document}